\documentclass[11pt]{article}

\usepackage{verbatim}
\usepackage{amsmath}
\usepackage{amsfonts}
\usepackage{amssymb}
\usepackage{bbm}
\usepackage{latexsym}
\usepackage{nicefrac}
\usepackage{graphics}
\usepackage{lscape} 
\usepackage{epsfig}
\usepackage{color}
\usepackage{multirow}
\usepackage{amscd}
\usepackage{cite}

\long\def\symbolfootnote[#1]#2{\begingroup%
\def\thefootnote{\fnsymbol{footnote}}\footnote[#1]{#2}\endgroup}
\usepackage{times}
\usepackage{graphicx}

\usepackage[small,sc]{caption}

\renewcommand{\d}{\mathrm{d}}

\DeclareMathSymbol{\mg}{\mathrel}{symbols}{"1D}

\newcommand{\Lm}{\mathsf{L}}
\newcommand{\Rm}{\mathsf{R}}

\newcommand{\ga}{\alpha}
\newcommand{\gb}{\beta}
\renewcommand{\gg}{\gamma}
\newcommand{\gd}{\delta}

\newcommand{\gx}{\xi}
\newcommand{\gm}{\mu}
\newcommand{\gn}{\nu}

\newcommand{\gl}{\lambda}
\newcommand{\gr}{\rho}

\newcommand{\gs}{\sigma}
\newcommand{\gt}{\tau}

\newcommand{\gp}{\pi}
\newcommand{\gps}{\psi}
\newcommand{\get}{\eta}

\newcommand{\gD}{\Delta}

\newcommand{\gL}{\Lambda}

\newcommand{\gO}{\Omega}

\newcommand{\cR}{{\cal R}}

\newcommand{\cV}{{\cal V}}

\newcommand{\tN}{{\widetilde N}}

\newcommand{\tr}{\text{tr}}

\newcommand{\Id}{\text{\small 1}\hspace{-3.5pt}\text{1}}

\newcommand{\ra}{\rightarrow}

\newcommand{\der}{\partial}

\newcommand{\dsp}{\displaystyle}

\newcommand{\beq}{\begin{equation}}
\newcommand{\eeq}{\end{equation}}
\newcommand{\barr}{\begin{array}}
\newcommand{\earr}{\end{array}}
\newcommand{\equ}[1]{\begin{gather} #1 \end{gather}}
\newcommand{\equa}[1]{\begin{align} #1 \end{align}}
\newcommand{\items}[1]{\begin{itemize} #1 \end{itemize}}
\newcommand{\enums}[1]{\begin{enumerate} #1 \end{enumerate}}

\newcommand{\arry}[2]{\begin{array}{#1} #2 \end{array}}

\newcommand{\non}{\nonumber}
\newcommand{\sfrac}[2]{\mbox{$\frac{#1}{#2}$}}
\newcounter{oldcounter}

\newcommand{\bder}{\bar\partial}
\newcommand{\bA}{{\overline A}}

\newcommand{\bget}{{\bar\eta}}

\newcommand{\bgD}{{\overline\Delta}}

\newcommand{\Intr}{\mathbb{Z}}
\newcommand{\Cplx}{\mathbb{C}}
\newcommand{\Real}{\mathbb{R}}

\newcommand{\ba}[2]{\[\begin{array}{#2}\label{#1}}
\newcommand{\ea}{\end{array}\]}
\newcommand{\be}{\begin{equation}}
\newcommand{\ee}{\end{equation}}
\newcommand{\bea}{\begin{eqnarray}}
\newcommand{\eea}{\end{eqnarray}}

\newcommand{\rep}[1]{\mathbf{#1}}
\newcommand{\crep}[1]{\overline{\rep{#1}}}

\newcommand{\sm}{{\,\mbox{-}}}

\def\er{\mathbb{R}}
\def\en{\mathbb{N}}
\def\zi{\mathbb{Z}}
\def\ci{\mathbb{C}}
\def\di{\mathrm{d}}

\def\slr{SL(2,\mathbb{R})}

\def\p{\partial}
\def\pb{\bar{\partial}}

\def\Id{{\rm 1\kern-.28em I}}
\def\ds{\displaystyle}
\newcommand{\oao}[2]{{#1\atopwithdelims[]#2}}

\def\ee{\mathrm{e}}
\def\d{{\partial}}

\begin{document}

\thispagestyle{empty}

\begin{flushright}
CPHT-RR076.0713 \\ 
LMU-ASC 54/13 \\
%Version: \today  
%\\
\end{flushright}
\vskip 2 cm
\begin{center}
{\Large {\bf Heterotic warped Eguchi--Hanson spectra with five--branes and line bundles} 
}
\\[0pt]

\bigskip
\bigskip {\large
{\bf Luca Carlevaro\footnote{
E-mail: Luca.Carlevaro@cpht.polytechnique.fr},
\bigskip }}\\[0pt]
\vspace{0.23cm}
{\it 
Centre de Physique Th\'eorique, 
Ecole Polytechnique, 91128 Palaiseau, France
 \\} 
\bigskip {\large
{\bf Stefan Groot Nibbelink\footnote{
E-mail: Groot.Nibbelink@physik.uni-muenchen.de},
\bigskip }}\\[0pt]
\vspace{0.23cm}
{\it 
Arnold Sommerfeld Center for Theoretical Physics, Ludwig-Maximilians-Universit\"at M\"unchen, 
\\
Theresienstrasse 37, 80333 M\"unchen, Germany
 \\} 
\bigskip
\end{center}

\subsection*{\centering Abstract}

We consider heterotic strings on a warped Eguchi--Hanson space with five--brane and line bundle gauge fluxes. 
The heterotic string admits an exact CFT description in terms of an asymmetrically gauged 
$SU(2)\times SL(2,\mathbbm{R})$ % SU(2)xSL(2,R) 
WZW model, in a specific double scaling limit in which the blow--up radius and the string scale are sent to zero simultaneously. 
This allows us to compute the perturbative 6D spectra for these models in two independent fashions: 
i) Within the supergravity approximation we employ a representation dependent index;
ii) In the double scaling limit we determine all marginal vertex operators of the coset CFT. 
To achieve agreement between the supergravity and the CFT spectra, 
we conjecture that the untwisted and the twisted CFT states correspond to the same set of hyper multiplets in supergravity. 
This is in a similar spirit as a conjectured duality between asymptotically linear dilaton CFTs and little string theory living on NS--five--branes. 
As the five--brane charge is non--vanishing, heterotic (anti--)five--branes have to be added in order to cancel irreducible gauge anomalies. 
The local spectra can be combined in such a way that supersymmetry is preserved on the compact resolved 
$T^4/\mathbbm{Z}_2$ % T^4/Z_2
orbifold by choosing the local gauge fluxes appropriately.

\newpage 
\setcounter{page}{1}

\section{Introduction}
\label{sc:intro}

Progress in string theory has often been driven by identifying backgrounds on which strings can be quantized exactly as a Conformal Field Theory (CFT). Flat orbifold CFTs provide some of the most easy examples \cite{Dixon:1985jw,Dixon:1986jc,Ibanez:1987sn,Font:1988tp,Ibanez:1988pj}. However, generically in the process of deforming the theory to live on a smooth Calabi--Yau geometry, typically the exact string quantization is lost.

It is therefore quite remarkable that in \cite{Carlevaro:2008qf} it was found that heterotic strings on a warped Eguchi--Hanson space in a specific double scaling limit can still be quantized exactly. In this double scaling limit both the blow--up radius and the string scale are send to zero simultaneously while their ratio is kept fixed so that the coupling remains finite everywhere. The resulting CFT builds on Wess--Zumino--Witten (WZW) coset theories based on an asymmetric gauging of $SU(2)_k\times SL(2,\Real)_{k'}$. (See e.g.\ the textbook of \cite{DiFrancesco:1997nk} and Ref.\ \cite{Israel:2004xj} for more concise descriptions.) We implement the line bundle in the coset CFT differently with respect to~\cite{Carlevaro:2008qf}, which allows us to exactly recover the integrated Bianchi identity in supergravity from the anomaly cancellation condition on the worldsheet theory. In contrast, the condition on the first Chern class of the line bundle, guaranteeing its stability, is now stronger than the usual K--theory condition arising heterotic supergravity. This conditions is nonetheless consistent with CFT results for heterotic strings on the non--compact $\ci^2/\zi_2$ orbifold~\cite{GrootNibbelink:2003zj}.

The exact CFT description of heterotic warped Eguchi--Hanson backgrounds involves both continuous and discrete representations
of the coset $\slr/U(1)$: States falling in continuous representations are generically massive non--localized bulk states concentrated away from the blown--up two--cycle of the Eguchi--Hanson geometry, while discrete representations are localized in its vicinity. In particular, the presence of discrete representations brings about the partial breaking of the gauge group. This exact coset CFT description allows us to determine the full partition function for continuous representations for all consistent gauge bundles. In contrast the partition function for discrete representations has to be worked out on a case by case basis, since it partially breaks gauge invariance by involving an $SO(32)_1/U(1)$ coset, where the gauging of the Abelian factor depends on the line bundle chosen. Nevertheless, we can derive the full spectrum of massless states for an arbitrary line bundle in the asymptotic limit for the radial coordinate of $\slr/U(1)$ by constructing the corresponding marginal vertex operators in the CFT. In particular,
bosonic states giving the CFT description of hyper multiplets in supergravity are shown to be in discrete representations of $\slr/U(1)$, while massless gauge multiplets are generically non-normalizable. We thus develop a systematic method to compute CFT spectra for the near--horizon limit of heterotic warped Eguchi--Hanson backgrounds in this paper, based on the new way of implementing the line bundle in the partition function. These computations systematize and extend those performed in \cite{Carlevaro:2008qf} and \cite{Carlevaro:2012rz} for very specific line bundle backgrounds.

A central aim of this paper is to compare the resulting massless CFT spectra with the zero mode spectra determined within the supergravity approximation. To do so, we determine a representation dependent multiplicity operator (index) \cite{Nibbelink:2008tv,Nibbelink:2007pn,GrootNibbelink:2007ew,Nibbelink:2007rd} for unwarped and warped Eguchi--Hanson spaces supporting line bundle backgrounds. Consistency of these spectra is confirmed by showing that their irreducible gauge anomalies can be cancelled by (anti--)five--brane contributions \cite{Aldazabal:1997wi,Honecker:2006dt}. As is well known anomaly cancellation in six dimension is extremely restrictive \cite{Green:1984bx,Erler:1993zy,Avramis:2005hc}, hence provides a stringent consistency check of the perturbative and five--brane spectra we determined. As the CFT can be formulated as a $\Intr_2$ orbifold,
the CFT spectra fall into two classes: untwisted and twisted. We show that the untwisted CFT and the supergravity spectra can be matched identically while the twisted states seem to be redundant. By using proposed dualities between asymptotically linear dilaton CFTs to little string theories living on \textsc{ns} five-brane configuration~\cite{Aharony:2004xn},  we conjecture that the twisted  CFT spectra describe to the same set of hyper multiplets as the untwisted ones. Since the corresponding untwisted and twisted operators generate marginal deformations of the CFT, we argue in particular that 
the couplings of the latter are not independent so that they must be turned on simultaneously. We also show that the corresponding twisted vertex operators are actually non-perturbative (in $\alpha'$) and thus can be seen in the supergravity limit as non-perturbative completion 
for the untwisted states.

The spectra on the warped Eguchi--Hanson space obtained in this paper can be used to describe the spectrum on the blown up $T^4/\Intr_2$ orbifold with different gauge fluxes localized near different resolved orbifold singularities. In a specific model, studied before e.g.\ in Refs.\ \cite{Honecker:2006qz,Nibbelink:2007rd}, the local Bianchi identities are violated near all fixed points; only the global Bianchi identity is fulfilled. Hence locally near all the resolved singularities there are either five--brane or anti--five--brane contributions. We show that they precisely compensate each other leaving only the perturbative spectrum on a compact K3 as expected.

\subsection*{Paper overview}

Section~\ref{sc:geom} describes the Eguchi--Hanson geometry and its warped extension. Subsection~\ref{sc:leftinv} gives the defining line element and the characteristic two--form it supports. In the next Subsection its characteristic topological data is obtained using toric geometry. In Subsection~\ref{sc:warped} the five--brane charged is introduce as a measure of the amount of warping the Eguchi--Hanson space experiences. The warped geometry admits two special limits discussed in the final part of this Section. One of which, the double scaling limit, is of prime interest in the CFT discussion in this work. 

In Section~\ref{sc:models} we investigate heterotic supergravity on the warped Eguchi--Hanson space. In its first Subsection we characterize line bundle backgrounds on this geometry. In Subsection~\ref{sc:spectra} we determine the charged spectra on the (warped) Eguchi--Hanson space. Using these results we investigate how either five--branes or anti--five--branes are involved in the cancellation of irreducible six--dimensional gauge anomalies. 

Section~\ref{sc:CFTwarpedEH} discusses a CFT realization of the heterotic string on a warped Eguchi--Hanson geometry in the double scaling limit. Subsection~\ref{sc:AsymGauging} shows that the resulting worldsheet theory can be described as a asymmetrically gauged WZW model. In Subsection~\ref{sc:WarpedEHorbifold} we show that in a specific gauge a residual $\Intr_2$ orbifold identification remains. This allows us to obtain the partition function for the continuous representations of this CFT. In Subsections~\ref{sec:cftmodels} and~\ref{sec:gaugemu} we determine the massless hyper multiplet and gauge multiplet spectra, respectively, in the CFT description, by determining the corresponding marginal vertex operators characterized in Subsection~\ref{sc:MasslessCFTspec}.

In Section~\ref{sc:compar} we compare the supergravity and CFT spectra. In order to achieve full agreement we conjecture that the twisted states are related to the untwisted ones, as the latter can be matched one--to--one to the supergravity states. To support this conjecture we review some results on marginal deformations of asymptotically linear dilaton CFTs and the conjectured duality between these CFTs and little string theories on NS five--brane configurations, in Subsections~\ref{sc:deformations} and \ref{sc:AdSCFT}, respectively. 

Section~\ref{sc:compactmodels} investigates how the results obtained in the previous Sections can be incorporated in a global compact description. Section~\ref{sc:concl} summarizes our main findings. Some useful trace formulae in the analysis of anomalies  in six dimensions are collected in Appendix~\ref{sc:traces}. In Appendix~\ref{sc:thetachar} we give some details of theta--functions and characters used to represent the partition function in Section~\ref{sc:WarpedEHorbifold}.

\newpage 
\section{Descriptions of (warped) Eguchi--Hanson backgrounds} 
\label{sc:geom}

In this section we describe the standard Eguchi--Hanson space both by giving its explicit metric as well as using toric geometry. After that we consider its warped generalization. This warped Eguchi--Hanson space, in addition to the conventional orbifold limit, admits a special double scaling limit which zooms in to the near--horizon region of the warped geometry.

\subsection{Eguchi--Hanson geometry}
\label{sc:leftinv} 

% EH metric 

The metric of the Eguchi--Hanson space $\mathcal{M}_\textsc{eh}$ can be represented as \cite{Eguchi:1978xp,Eguchi:1980jx}
\begin{equation}\label{metric}
\di s^2_\textsc{eh} = 
%\Big( 1-\frac{a^4}{r^4} \Big)^{-1} 
g^{-2}(r)\, \di r^2
+
\frac{r^2}{4} \Big[
(\sigma^\mathsf{L}_1)^2 + (\sigma^\mathsf{L}_2)^2 +
%\Big(1-\frac{a^4}{r^4}\Big)
g^2(r)
 (\sigma^\mathsf{L}_3 )^2 
\Big]~,
\end{equation}
with $g(r) = \sqrt{1-\frac{a^4}{r^4}}$ for $r > a$, in terms of the $SU(2)$ left--invariant one--forms:
\begin{equation}\label{1f}
\begin{array}{lll}
{\ds \sigma^\mathsf{L}_1 = \sin\psi \, \di\theta - \cos\psi \sin\theta\, \di\phi }~, &\quad
 {\ds \sigma^\mathsf{L}_2 = -\cos\psi \, \di\theta - \sin\psi \sin\theta\, \di\phi }~, & \quad
 {\ds \sigma^\mathsf{L}_3 =  \di\psi + \cos\theta \, \di\phi }~.  
\end{array}
\end{equation}
The original periodicities of the Euler angles parameterizing the $SU(2)$ group manifold read: $\theta\in [0,\pi]$, $\phi\in [0,2\pi]$ and $\psi\in [0,4\pi]$. In the description of the smooth Eguchi--Hanson space the periodicity of the $\psi$ coordinate is modified and runs over half of its original span $\psi \in [0,2\pi]$: The extra $\Intr_2$ orbifold action $\psi \ra \psi + 2\pi$ is necessary to eliminate the bolt singularity at $r=a$; ensuring that the Eguchi--Hanson geometry is smooth for any $a > 0$. The metric defines a self--dual K\"ahler form:
\begin{equation}
J_\textsc{eh}= {e}^0\wedge {e}^3+{e}^1\wedge {e}^2~, 
\qquad 
*_\textsc{eh} J_\textsc{eh}  = J_\textsc{eh}~, 
\end{equation}
in terms of the associated vielbein one--forms $e^a$ with frame indices $a=0,\ldots, 3$:
\begin{equation}\label{vier}
{e}^0 = \frac{1}{g(r)}\, \di r~, 
\qquad 
{e}^a = \frac{r}{2}\,\sigma_a^{\textsc{l}}~, 
a =1,2~, 
\quad \text{and}\quad 
e^3 = \frac{r}{2}\,g(r)\, \sigma_3^{\textsc{l}}~.
\end{equation}

% P^1 cycle and associated 2-form 

There is a two--cycle with the topology of $\mathbb{P}^1$ located within the Eguchi--Hanson space at $r =a$. It is given geometrically as a non--vanishing two--sphere 
\begin{equation} \label{KahlerP1} 
\di s^2_{\mathbb{P}^1} = \frac{a^2}{4}\big(\di\theta^2+ \sin^2\theta\,\di\phi^2\big)~, 
\qquad 
 J_{\mathbb{P}^1} = - \tfrac14 \sigma^\mathsf{L}_1 \wedge  \sigma^\mathsf{L}_2 = \tfrac14 \sin\theta \, \di\theta \wedge \di\phi \,.
\end{equation}
Its Poincar\'e dual two--form on the Eguchi--Hanson geometry has the {\it local} description:
\begin{equation}\label{oeh}
\mathcal{F}_\textsc{eh} 
=  - \frac{a^2}{4\pi}\,\di \Big(\frac{\sigma^\textsc{l}_3}{r^2} \Big) 
= -\frac i2 \left( \frac{a}{r}\right)^2 \left[ \sin\theta\, \di \theta \wedge \di \phi + \frac2r\, \di r \wedge
 \sigma^\mathsf{L}_3 \right]~. 
\end{equation}
This gauge field strength can be expressed in terms of the vielbein one--forms as 
\begin{equation}\label{oeh-viel}
\mathcal{F}_\textsc{eh} 
= \frac{a^2}{\pi r^4}\Big({e}^0\wedge{e}^3-{e}^1\wedge {e}^2\big)~, 
\qquad 
*_\textsc{eh} \mathcal{F}_\textsc{eh} = - \mathcal{F}_\textsc{eh}~, 
\quad 
J_\textsc{eh} \wedge \mathcal{F}_\textsc{eh} = 0~,  
\end{equation}
showing that $\mathcal{F}_\textsc{eh}$ is {\it anti--self--dual} and orthogonal to the K\"ahler form of the Eguchi--Hanson space. This two--form can be interpreted as the field strength associated to an Abelian gauge background supported on the Eguchi--Hanson geometry with the following integral properties: 
\begin{equation}\label{gaugeintegrals}
 \int_{\mathbb{P}^1} \mathcal{F}_\textsc{eh} = 1\,, 
 \quad \text{and } \quad 
 \int_{\mathcal{M}_\textsc{eh}} \mathcal{F}_\textsc{eh} \wedge \mathcal{F}_\textsc{eh} = - \frac 12\,.
\end{equation}
Consequently, the second cohomology $H^2(\mathcal{M}_\textsc{eh})$ is spanned by a single generator $[\mathcal{F}_\textsc{eh}]$; an unique representative is given by the harmonic and anti--selfdual two-form~\eqref{oeh-viel}.

\subsection{Eguchi--Hanson geometry as a toric variety}
\label{sc:toricC2Z2} 

It is also possible to describe the Eguchi--Hanson geometry using toric geometry, see e.g.\ \cite{Nibbelink:2007pn,Nibbelink:2008qf}. In the toric description the Eguchi--Hanson background is described in terms of three complex coordinates$(z_1,z_2,x)$ defining the toric variety, 
\equ{
(z_1,z_2,x) \sim (\gl\, z_1,\gl\, z_2, \gl^{-2}\, x) \in \frac{\mathbbm{C}^3-\mathbbm{F}}{\mathbbm{C}^*}~, 
}
where $\gl \in \mathbbm{C}^*$ and the exclusion set $\mathbbm{F}$ is given $\mathbbm{F} := \{z_1=z_2 = 0\}$. The ordinary divisors, 
$D_1 := \{z_1 = 0\}$, $D_2 := \{z_2 = 0\}$,
and exceptional divisor, $E := \{x = 0\}$, are subject to the linear equivalence relations 
\equ{
2\, D_1+ E \sim 0~, 
\qquad 
2\, D_2+ E \sim 0~. 
\label{linequiv}
}
The unambiguously defined intersection numbers are
\equ{
D_1E = D_2E = 1~, 
\qquad 
E^2 = -2~, 
}
since they all involve the compact divisor $E$. The total Chern class is given via the splitting principle $c = (1+D_1)(1+D_2) (1+E)$. Expanding this out using the linear equivalence relations \eqref{linequiv} we find 
\equ{
c_1 = D_1+D_2+E \sim 0~, 
\qquad 
c_2= D_1D_2 + D_1E + D_2 E \sim  - \frac{3}{4}\, E^2 = \frac 32~. 
}
The first relation says that this toric variety is Calabi--Yau. The second Chern class agrees with an explicit computation using the Eguchi--Hanson metric \eqref{metric}. Finally, we note that the class $[\mathcal{F}_\textsc{eh}] \sim - \sfrac 12\, E$.

\subsection{Warped Eguchi--Hanson backgrounds}
\label{sc:warped}

The Eguchi--Hanson geometry can be warped \cite{Carlevaro:2008qf} by including a scalar factor $H(r)$ in the metric 
\begin{equation}\label{metric2}
\di s^2_{\textsc{eh}_W} = H(r) \di s^2_\textsc{eh}~,
\qquad 
H(r) =   1+ \frac{4\alpha' \mathcal{Q}_5 }{r^2}~, 
\end{equation}
with the $\di s^2_\textsc{eh}$ given in \eqref{metric}. This defines a supergravity background provided that we include a varying dilaton $\Phi(r)$ and a non--trivial three--form $\mathcal{H}$ given by 
\equ{ 
\mathrm{e}^{2 \Phi(r)}  = g_s^2\, H(r) 
\qquad 
\mathcal{H} = -H(r) \ast_{\textsc{eh}} \di H(r) = 8\alpha' \mathcal{Q}_5   \left(  1- \frac{ a^4}{r^4} \right)
 \text{Vol}(S^3)~,   
\label{solH}
}
where volume of the three--sphere   
\(
\text{Vol}(S^3) 
= \tfrac18\,  \sigma^\textsc{l}_1\wedge  \sigma^\textsc{l}_2\wedge  \sigma^\textsc{l}_3
= \tfrac18 \,\di(\cos\theta)\wedge\di\phi\wedge\di\psi
\) 
is given in terms of the Euler angles. 

% Torsion connections

The presence of the three--form $\mathcal{H}$ leads to torsional spin--connections 
\begin{equation}
\Omega_{\pm\phantom{a}b}^{\phantom{\pm}a} 
= \omega_{\phantom{a}b}^{a} 
\pm \tfrac12 \mathcal{H}_{\phantom{a}b}^{a}~. 
\end{equation}
where $\omega^a{}_b$ denotes the spin--connection including warping and $a,b$ are four dimensional frame indices.  Computing the second Chern class using the associated curvatures $\cR_\pm = \di \gO_\pm + \gO_\pm^2$ gives the same answer as in the unwarped case: 
\equ{ 
c_2(\cR_\pm) 
= - \frac 12\, \int \tr \Big( \frac{\cR_\pm}{2\pi}\Big)^2 
= \frac 32~. 
\label{c2warped} 
}
%

% The NS5-brane charge

This background describes an Eguchi--Hanson geometry supporting a non--vanishing three--form flux from the torsion $\mathcal{H}$ given by  
\begin{equation}\label{largeQ5}
2\mathcal{Q}_5 = \frac{-1}{4\pi^2 \alpha'} \int_{\partial \mathcal{M}_\textsc{eh}}\mathcal{H}~. 
\end{equation}
In blow down this can be interpreted as the brane charge of a stack of five--branes located at the orbifold singularity. In blow--up these five--branes are no longer visible as the radial coordinate $r$ starts at $r = a > 0$, while according to \eqref{solH} the five--branes remain positioned at $r=0$, hence only the effects of the torsion can be felt.

\subsection{Limits of the warped Eguchi--Hanson space}
\label{sc:limit} 

The warped Eguchi--Hanson geometry admits the following two interesting limits: 
\enums{ 
\item[i)] In the {\it blow-down limit} the parameter $a$ is taken to $a \ra 0$ and we recover a warped $\Cplx^2/\Intr_2$ background with constant three-form $\mathcal{H}$, which indicates the presence of \textsc{ns} five-branes opening a throat at $r=0$. Without warping the heterotic string can be exactly quantized in this limit and e.g.\ its spectrum can be determined. The connection between the exact orbifold CFT and the supergravity in blow--up was studied in \cite{Nibbelink:2007rd,Nibbelink:2008tv}. 
\item[ii)] The {\it near--horizon limit} is defined as the following double scaling limit~\cite{Carlevaro:2008qf}  
\begin{equation}\label{DSL}
a \ra 0~, 
\quad\text{and}\quad 
g_s \to 0~ 
\quad 
\text{keeping} 
\quad 
\lambda = \frac{g_s \sqrt{\alpha'}}{a} \quad \text{fixed}~.
\end{equation}
}
In the double--scaling limit the constant piece in the warp--factor \eqref{metric2} can be neglected against the $1/r^2$ contribution, assuming $\mathcal{Q}_5 > 0$. This will always be our assumption for the resolved geometry. By taking this limit, after scaling out the blow--up parameter $a$ from the new radial variable, i.e.\ $\cosh \rho = (r/a)^2$, one zooms into the region near the blown up two--cycle:
\begin{equation}\label{DSgeom}
\di s^2_\textsc{nh}  = 
    \alpha' \mathcal{Q}_5\Big[ {\rm d}\rho^2 + (\sigma^\mathsf{L}_1)^2+
(\sigma^\mathsf{L}_2)^2 + \tanh^2 \rho \, (\sigma^\mathsf{L}_3 )^2\Big]\,.
\end{equation}
In addition, the NSNS--three--form and the dilaton are also affected by the double--scaling limit:
\begin{subequations}\label{eqDSL}
\begin{align}
\mathcal{H}_\textsc{nh} & = -8\alpha' \mathcal{Q}_5\,\tanh^2 \rho\, \text{Vol}(S^3)~, \qquad 
\mathrm{e}^{2 \Phi_\textsc{nh}(\rho)}  =  \frac{4\lambda^2 \mathcal{Q}_5}{\cosh \rho}~, 
\label{dilateq}\\
\mathcal{F}_\textsc{nh} & =    -\frac{1}{2\cosh\rho}\big(\tanh\rho\,\di\rho\wedge\sigma^\mathsf{L}_3 - \sigma^\mathsf{L}_1\wedge\sigma^\mathsf{L}_2\big)~. \label{Feq}
\end{align}
\label{sugrasol}
\end{subequations}
The Abelian gauge field does not feel this limit; it is only written in terms of the new radial variable $\gr$. 

This means that the warped Eguchi--Hanson space stays resolved,  but the asymptotical properties of the metric at infinity have changed: the boundary of the warped Eguchi--Hanson space has been decoupled. This is reflected in the fact that the second Chern classes for torsional connections do not agree in the near--horizon limit: 
\equ{
c_2(\cR_{\textsc{nh}\pm}) 
= - \frac 12\, \int\tr \Big( \frac{\cR_{\textsc{nh}\pm}}{2\pi}\Big)^2 
= \frac 32 \pm \frac 12~. 
\label{c2nearh} 
}
The change of $\sfrac12$ unit with respect to \eqref{c2warped} is a consequence of the near--horizon limit which the boundary at infinity has been cut off.

\section{Heterotic supergravity on (warped) Eguchi--Hanson spaces}
\label{sc:models}

A heterotic supergravity background is defined by the smooth warped Eguchi--Hanson geometry $\mathcal{M}_\textsc{eh}$ supporting a gauge field strength $\mathcal{F}$. In this paper we only consider an $SO(32)$ gauge field (the extension to the $E_8\times E_8'$ case is straightforward).  For consistency such a background has to satisfy the Bianchi identity~\cite{Strominger:1986uh}:
\begin{equation}\label{bi}
- \sfrac 1{\alpha'} \, 
\di\mathcal{H} = \textsc{tr} \mathcal{F}\wedge \mathcal{F} - \textsc{tr}\,\mathcal{R}(\Omega_-)\wedge
\mathcal{R}(\Omega_-)~.
\end{equation}
 Here the traces $\textsc{tr}$ are defined in the vector representations of the gauge group $SO(32)$ and the Lorentz group $SO(1,9)$, respectively. If the three--form \eqref{solH} is not closed, the Bianchi identity \eqref{bi} requires a non--standard embedding of the Lorentz connection into the gauge connection. For particular local solutions to this identity see e.g.~\cite{Fu:2006vj,Becker:2006et}.

\subsection{Line bundle embeddings}
\label{sc:general} 
% Abelian background embedding 

Since the study of non--standard embedding bundles is rather challenging, 
we focus here only on line bundle gauge fluxes. These Abelian backgrounds can be studied systematically. In general the embedding of the direct sum of line bundles on the Cartan subalgebra of $SO(32)$ can be described by
\begin{equation} \label{gaugebund}
 \frac{\mathcal{F}_{\mathbf{Q}}}{2\pi} = \frac{\mathcal{F}_\textsc{eh}}{2\pi}\, \sum_{I=1}^{16}Q_{I} \,\textsf{H}^I \,.   
\end{equation}
The two--form $\mathcal{F}_\textsc{eh}$, defined in \eqref{oeh}, is supported on the warped Eguchi--Hanson space. The anti--Hermitean Cartan generators $\textsf{H}^a\in \mathfrak{h}(SO(32))$ are  normalized such that $\textsc{tr}\,\textsf{H}^I\textsf{H}^J=-2\delta^{IJ}$. The Abelian gauge background is characterized by the {\it line bundle vector} $\mathbf{Q} = (Q_1,\ldots, Q_{16})$ and is well--defined when the following conditions are met: 
\enums{ 
\item[i)] A Dirac quantization condition on the gauge instanton charge :
\equ{ 
\mathbf{Q}\in \gL_{16}~, 
\qquad 
\gL_{16} =  \zi^{16} \oplus \Big(  \zi^{16} + \tfrac12\, \mathbf{e}_{16} \Big)~, 
 \label{dirac}
 } 
where $\mathbf{e}_{16} = (1,\ldots, 1)$ is the vector with sixteen entries equal to one. It has been shown that the two possible choices of integral or half-integral line bundle vectors correspond to the distinction between models characterized by gauge bundles $V$ without or without vector structure~\cite{Berkooz:1996iz}.

\item[ii)]  A stability condition for the gauge bundle $V$ requires that the first Chern class of the line bundle $\mathcal{L}$ to be 
in the second even integral cohomology class of the Eguchi--Hanson space~\cite{Witten:1985mj,Freed:1986zx,Blumenhagen:2005ga}:
\equ{ 
{\ds  c_1(\mathcal{L}) \in H^2(\mathcal{M}_\textsc{eh},2\zi) \quad \Rightarrow  \qquad 
\tfrac 12\, \mathbf{e}_{16} \cdot \mathbf{Q} = 
\tfrac 12\, \sum_{I} Q_I \equiv 0 \text{ mod } 1 \,.}
\label{bs}
}
This allows for spinorial representations of $V$ to appear at the massive level.
}
The metric~(\ref{metric2}), the NSNS--three--form~(\ref{solH}), the dilaton~\eqref{dilateq} and the line bundle background~(\ref{gaugebund}) 
satisfy the Bianchi identity~(\ref{bi}) in cohomology only in the large $\mathcal{Q}_5$ limit, in which $c_2(\mathcal{R}_-)$ can be neglected with respect to the line bundle contribution. For finite values of the five--brane charge, the conformal factor $H(r)$ and the dilaton generically receive corrections in $1/\mathcal{Q}_5$ (see for instance~\cite{Fu:2008ga,Dasgupta:1999ss}). Consistent heterotic backgrounds nevertheless only correspond to tadpole free models, for which the defining line bundle solves the integrated Bianchi identity~(\ref{bi}).
 
\subsection{Perturbative charged spectrum computation} 
\label{sc:spectra}

We compute the perturbative charged spectrum, i.e.\ part of the spectrum that in heterotic supergravity can be obtained as zero modes of the Dirac operator of the gaugino. This spectrum can be computed by employing the following multiplicity operator \cite{Nibbelink:2007rd}
\equ{
\textsc{N}_\mathbf{Q} = - \int_{\mathcal{M}_\textsc{EH}} 
\Big\{ 
\frac 12\, \Big( \frac{\mathcal{F}_\mathbf{Q}}{2\pi}\Big)^2
+ \frac1{24}\, \text{tr} \Big( \frac{\mathcal{R}}{2\pi}\Big)^2 
\Big\}~. 
\label{MultOp}
}
This operator can be thought of as a generalization of a representation dependent index. The integrals can be evaluated using the results of Section \ref{sc:geom}.

Given that the second Chern class in the near--horizon limit is different from that of the non--warped Eguchi--Hanson space and depends on which torsion connection is employed, it is useful to rewrite the multplicity operator \eqref{MultOp} in a form which does not explicitly depend on $c_2$. Integrating the Bianchi identity \eqref{bi} leads to a relation between the second Chern class $c_2$, the five--brane charge $\mathcal{Q}_5$ and the line bundle vector $\mathbf{Q}$: 
\equ{
- \mathcal{Q}_5 =2\, c_2 -  \tfrac12\, \mathbf{Q}^2~. 
\label{intBI}
}
By using this and computing the integrals in \eqref{MultOp} we obtain an explicit formula for the multiplicity operator
\equ{
\textsc{N}_\mathbf{Q} 
= \frac 14\, \textsc{H}_\mathbf{Q}^2 - \frac 1{12}\, c_2 
= \frac 14\, \textsc{H}_\mathbf{Q}^2 
- \frac 1{48}\, \mathbf{Q}^2 + \frac 1{24}\, \mathcal{Q}_5~.  
\label{MultOp_eval}
}

To evaluate the multiplicity operator on the states that an $SO(32)$ gaugino consists of, we represent the $SO(32)$ roots as sixteen component vectors with two entries $\pm 1$ and the rest zero, i.e.\ 
$\mathbf{P} = (\underline{\pm 1^2, 0^{14}})$. (Here the powers denote the number of times an entry appears and the underline means all possible permutations.) Consequently we have 
\(
[\textsc{H}_\mathbf{Q}, E_\mathbf{P}]
= \mathbf{Q} \cdot \mathbf{P}\, E_\mathbf{P}
\) 
for the $SO(32)$ generators $E_\mathbf{P}$ associated with the roots $\mathbf{P}$. The normalization of the multiplicity operator \eqref{MultOp} is chosen such that it counts the number of hyper multiplets in a complex representation. This means that a half--hyper multiplet, i.e.\ a hyper multiplet on which an reality condition is enforced, gets half the multiplicity of that of a full hyper multiplet. Moreover, the sign of the multiplicity operator distinguishes between hyper multiplets and vector multiplets. However, given that the vector multiplets take values in the adjoint representation, we need to include an additional factor of a half for them in the spectrum. 

% Sugra spectra 
%
\begin{table}
\[
\renewcommand{\arraystretch}{1.5} 
\begin{array}{|c|c|r|}\hline 
\multicolumn{2}{|c}{\text{Gauge group:}~ U(N_1)\times \ldots \times U(N_n) \times SO(2N)} & \multicolumn{1}{|c|}{\textsc{N}_\mathbf{Q}(\mathbf{P})} 
\\
\text{States} &  SO(32)\text{--Roots:}~~\mathbf{P}= &
 (2\,c_2 =\sfrac 12\, \mathbf{Q}^2 - \mathcal{Q}_5)
\\\hline\hline 
(\ldots \rep{1}, \rep{Ad}_{U(N_i)}, \rep{1} \ldots ; \rep{1})  & (\ldots 0, \underline{1,\sm1,0^{N_i-2}}, 0 \dots;0^N) & - \frac 1{12}\, c_2
\\
(\ldots \rep{1};\rep{Ad}_{SO(2N)})  & (0 \ldots 0,  
\underline{\pm1^2,0^{N-2}}) &  - \frac 1{12}\, c_2
\\ \hline 
(\ldots\rep{1}, [\rep{N_i}]_2, \rep{1},\ldots ;\rep{1})  & (\ldots 0, \underline{1^2,0^{N_i-2}},0\ldots;  
0^{N}) &  p_i^2 - \frac 1{12}\, c_2
\\
(\ldots\rep{1},\rep{N_i}, \rep{1}\ldots; \rep{2N}) & (\ldots 0, \underline{1,0^{N_i-1}},0 \ldots; 
\underline{\pm 1, 0^{N-1}}) &  \tfrac 14\, p_i^2 - \frac 1{12}\, c_2
\\
(\ldots \rep{1},\rep{N_i}, \rep{1}\ldots\rep{1}, \rep{N_j},\rep{1}\ldots; \rep{1}) & (\ldots 0,\underline{1,0^{N_i-1}}, 0\ldots0, \underline{1,0^{N_j-1}}, 0 \ldots;
0^{N}) &  \tfrac 14\, (p_i+p_j)^2 - \frac 1{12}\, c_2
\\ 
(\ldots \rep{1},\rep{N_i}, \rep{1}\ldots\rep{1}, \crep{N_j},\rep{1}\ldots; \rep{1})  & (\ldots 0,\underline{1,0^{N_i-1}}, 0\ldots0, \underline{\sm1,0^{N_j-1}}, 0 \ldots;
0^{N})  & \tfrac 14\, (p_i-p_j)^2 - \frac 1{12}\, c_2
\\\hline 
\end{array}
\]
\caption{Perturbative spectrum: The first two rows give the gauge multiplet representations while the rows below that give the representations and the multiplicities $N_\mathbf{Q}$ of the hyper multiplets. To shorten the notation we write $\ldots \rep{1}$ for $\rep{1},\ldots,\rep{1}$, etc.\ and with underlined entries we mean this set of entries and all possible permutations. $[\rep{N}]_2$ denotes the anti--symmetric rank--two tensor representation of $SU(N)$. 
\label{tb:pertspectrum}}
\end{table}

% Bundle Ansatz
%
As usual it is rather difficult to make statements about spectra in general, unless one uses a specific ansatz. An ansatz which can describe a large class of models is defined by a line bundle vector of the form 
\equ{\label{shiftAnsatz}
\mathbf{Q} = \big( 
p_1^{N_1},\ldots , p_n^{N_n}; 0^{N}
\big)~, 
\qquad 
 N_1 + \ldots + N_n + N  = 16~, 
}
Since a line bundle is actually determined by an equivalence class of such vectors under 
a shift by any element of the Narain lattice of $SO(32)$, we can take, without loss of generality: $p_1 > \ldots > p_n > 0$ with $p_i$ non--vanishing integers (or half integers when $N=0$). This line bundle background induces the gauge symmetry breaking 
\equ{ \label{GaugeGroup}
SO(32) \ra U(N_1) \times \ldots \times U(N_n) \times SO(2N)~. 
}
%
% Comment on the massive U(1)
%
This means that the gauge group contains $n$ Abelian $U(1)$ factors. 
The $U(1)_\mathbf{Q}$ gauge field associated to the Cartan direction set by $\mathbf{Q}$ becomes massive via the generalized Green--Schwarz mechanism, while the other perpendicular combinations remain massless.

% Spectrum overview
%
In Table \ref{tb:pertspectrum} we have listed the full perturbative spectrum computed using the multiplicity operator~\eqref{MultOp_eval}. We classify the states according to their $U(N_i)$, $U(N_j)$ 
(with the indices $1 \leq i < j \leq n$) and $SO(2N)$ representations. 
The charges of these states are determined by  $\textsc{H}_\mathbf{Q}$. In this table we only indicate the $\textsc{N}_\mathbf{Q}$ multiplicities of the various states. The first four rows of the Table give the gaugino states in the adjoints of the unbroken gauge groups; the states below that denote the six--dimensional hyper multiplet matter. To emphasize the systematics we do not make the breaking of $U(1)_\mathbf{Q}$ explicit in this table.

\subsection{Eguchi--Hanson models without five--brane charge}
\label{sc:anomalies}

% Anomalies and perturbative anomaly freedom
%
In general the perturbative spectrum suffers from irreducible pure gauge $SU(N_1)^4$, \ldots, $SU(N_2)^4$ and  $SO(2N)^4$--anomalies. Indeed, the resulting anomaly polynomial is proportional to 
\equ{ 
I_{8|pert} \supset 
\mathcal{Q}_5 \Big\{ 
\tr\, F_{SU(N_1)}^4 +\ldots +  \tr\, F_{SU(N_n)}^4 + 
\tfrac 12\, \textsc{tr}\, F_{SO(32-2N)}^4 
\Big\}~, 
\label{IrredAnoms} 
}
taking into account the additional factor of $\sfrac 12$ for the gauge multiplets mentioned above and using the trace identities given in Appendix \ref{sc:traces}. The traces are evaluated in the fundamental representations of the respective gauge groups.

When $\mathcal{Q}_5$ vanishes, i.e.\ when $\mathbf{Q}^2 = 4\, c_2$, the perturbative spectrum is free of irreducible anomalies. Such models  are sometimes referred to as perturbative models \cite{Aldazabal:1997wi}. For the non-warped Eguchi--Hanson background the perturbative models were classified in \cite{Nibbelink:2007rd} and are for completeness given in Table \ref{tb:pertEHmodels}. As was confirmed in Refs.\ \cite{Nibbelink:2007rd,Nibbelink:2008tv} these spectra are compatible with that of heterotic $\mathbbm{C}^2/\mathbbm{Z}_2$ orbifold models (for a classification is e.g.~\cite{Aldazabal:1997wi}) in blow--up.

In detail the interpretation of the multiplicities is as follows: The states with a positive factor of $\sfrac 1{16}$ are untwisted, i.e.\ ten--dimensional bulk states, given that $c_2 = \sfrac 32$ in this case. Since they are just the internal parts of the gauge fields they have an internal complex space index, hence always come in pairs. In total these bulk untwisted matter thus have a multiplicity $\sfrac 2{16}$. Localized twisted states have an integral multiplicity. Finally, the orbifold models often have an enhanced symmetry group, which is higgsed to gauge group in blow--up. In this process the massive gauginos pair up with a part of the twisted hyper multiplets, this ``eating of the Goldstone modes'' is reflected in Table \ref{tb:pertEHmodels} by the contributions $1-\sfrac{2}{16}$.

\subsection{Non--perturbative (anti--)five--brane states}
\label{sc:FiveBranes}

The models that involve non--vanishing five--brane charge, i.e.\ $\mathcal{Q}_5 \neq 0$, are called non--perturbative models. Since the (warped) Eguchi--Hanson space is non--compact, it is possible there is some anomaly inflow from infinity to ensure consistency. Hence in principle warped Eguchi--Hanson models with $\mathcal{Q}_5 \neq 0$ can be tolerated.

In blow down the Eguchi--Hanson geometry degenerates to an orbifold with five--branes located at $r=0$. The five--brane charge $\mathcal{Q}_5$ is then associated to this stack of five--branes: $-\mathcal{Q}_5$ is equal to twice the number of five--branes in the stack denoted by $\tilde N$. On the five--branes one expects an $Sp(\tN)$ gauge group to be present \cite{Witten:1995gx}. The non--perturbative spectrum is suggested by a conjectured S--duality with Type II orientifolds to be:
\equ{ 
\bigoplus\limits_i  (\ldots \rep{1}, \rep{N_i},\rep{1}\ldots; \rep{1}; \rep{2\tN}) 
+ \tfrac 12\, (\rep{1},\ldots, \rep{1}; \rep{32-2N}; \rep{2\tN}) 
+ (\rep{1},\ldots,\rep{1};\rep{1}; [\rep{2\tN}]_2)~. 
\label{nonpertspectrum}
}
Here $\rep{2\tN}$ and $[\rep{2\tN}]_2$ denote the fundamental and the anti--symmetric tensor representations of $Sp(\tN)$, respectively. Using the trace identities given in Appendix \ref{sc:traces} it is not difficult to check that the irreducible anomalies \eqref{IrredAnoms} of the perturbative spectrum are cancelled by the (anti--)five--brane  (anti--)hyper multiplet states in \eqref{nonpertspectrum}. In addition, the pure irreducible $Sp(\tN)$ anomalies vanish as well. These findings are in agreement with the results reported in \cite{Honecker:2006dt}. The sign of $\mathcal{Q}_5$ is significant: When $\mathcal{Q}_5> 0$ then the five--branes are actually anti-- five--branes that preserve supersymmetries with the opposite chirality as the perturbative heterotic supergravity in six dimensions and five--branes do. This means that if $\mathcal{Q}_5$ is negative (positive), the spectrum \eqref{nonpertspectrum} is the spectrum of (anti--)hyper multiplets with the same (opposite) chirality as the hyper multiplets of the perturbative sector.

When anti--hyper multiplets are required, this signals that six--dimensional supersymmetry is broken by the presence of anti--five--branes.  However, in blow--up, and therefore in particular in the near--horizon limit, we have seen that the five--branes are not part of the physical space, hence the anomaly inflow from infinity is the only way that the model can become consistent. Therefore, because in the orbifold point with $\mathcal{Q}_5 > 0$ supersymmetry is broken, it seems that this gives a dynamical mechanism that drives the system in blow--up.

\begin{table}
\[
\renewcommand{\arraystretch}{1.5} 
\begin{array}{|c|c|c|}\hline 
%\multicolumn{3}{|c|}
%{c_2 = 3/2~, \quad \mathbf{Q}^2=6~, \quad \mathcal{Q}_5=0}
%\\
\mathbf{Q} & \text{Gauge group} & \text{Representation} 
\\\hline\hline 
(1^6,0^{10}) & SU(6) \times SO(20) & 
%\frac 18 (6,20)_1 + \frac 78 (15,1)_2 
\frac 2{16}(\rep{6},\rep{20})_1 + ( 1 -\frac 2{16})\cdot (\rep{15},\rep{1})_2 
\\\hline 
(2,1^2,0^{13}) & SU(2) \times SO(26) & 
%\frac 1{8}  (2,26)_1  + \frac 18 (2,1)_1 + \frac 78 (1,1)_2 +  \frac 78 (1,26)_2 
%+ \frac{17}8 (2,1)_3
\frac 2{16}  (\rep{2},\rep{26})_1  + \frac 2{16} (\rep{2},\rep{1})_1 + ( 1 - \frac 2{16})\cdot (\rep{1},\rep{1})_2 
\\ &  & 
+ (1 -  \frac 2{16})\cdot (\rep{1},\rep{26})_2 
+ (2 + \frac2{16})\cdot (\rep{2},\rep{1})_3
\\\hline 
( -\tfrac 32, \tfrac 12^{15}) & SU(15) & 
%\frac 18 (105)_1 + \frac 18 (15)_1  + \frac 78 (15)_2 
\frac 2{16} (\rep{105})_1 + \frac 2{16} (\rep{15})_1  + (1-  \frac 2{16})\cdot (\rep{15})_2 
\\\hline \end{array}
\]
\caption{
The non--warped Eguchi--Hanson models without five--brane charge have been listed before in \cite{Nibbelink:2007rd} for the sake of completeness and facilitate comparisons we repeat them here. The $H_\mathbf{Q}$ charge is indicated as the subscript. In the text the explanation for the fractional multiplicities is given. 
\label{tb:pertEHmodels}}
\end{table}

\section{CFT  description of warped Eguchi--Hanson}
\label{sc:CFTwarpedEH}

%\blue{\items{\item Overview of what will be in this section}} 

\subsection{Heterotic CFT for warped Eguchi--Hanson in the double scaling limit}
\label{sc:AsymGauging}

% basic fields of the CFT before gauging

The warped Eguchi--Hanson resolution with a line bundle background can be described as a certain gauged WZW model~\cite{Carlevaro:2008qf}. The starting point is an $SU(2)_k \times SL(2,\Real)_{k'}$ WZW model with $(1,0)$--worldsheet supersymmetry. From the group elements $(g,g') \in SU(2)_k \times SL(2,\Real)_{k'}$, 
\equ{\label{SUSLgroupElmts}
 g= \ee^{\frac i2 \sigma_3 \phi} \ee^{\frac i2 \sigma_1 \theta}   \ee^{\frac i2 \sigma_3 \psi} \,, 
 \qquad
   g' = \ee^{\frac i2 \sigma_3 \phi'} \ee^{\frac 12 \sigma_1 \rho}   \ee^{\frac i2 \sigma_3 \psi'} \,,
}
one constructs left-- and right--moving (super--)currents: 
We have the following left--moving super--currents,  
\equ{ \arry{l}{
\big(J^\ga_\Lm=(k-2)\, (\der g\, g^{-1})^\alpha + \frac i2\varepsilon^{\alpha}_{\phantom{\alpha}\beta\gamma}\! :\!\gps_\Lm^\beta \gps_\Lm^\gamma\!: \,, \; \gps_\Lm^\ga \big)~,  
\\[2ex] 
\big(K^\alpha_\mathsf{L}=(k'+2)\, (\der g^{\prime}\, g^{\prime -1})^\alpha + 
\frac i2\varepsilon^{\alpha}_{\phantom{\alpha}\beta\gamma} \!:\!\psi^{\prime\beta} _\Lm\psi^{\prime \gamma}_\Lm\!: \,, \; \psi^{\prime\ga}_\Lm \big)~, 
}
}
forming the super--affine $\widehat{\mathfrak{su}}(2)_k$ 
and super--conformal $\widehat{\mathfrak{sl}}(2,\Real)_{k'}$ algebras, respectively. The bosonic currents,  
\equ{
j_\Rm^\alpha =  (k-2)\, (g^{-1}\bder g)^\alpha
\quad \text{and} \quad  
k^{\ga}_\Rm = (k'+2)\, (g^{\prime -1}\bder g' )^\ga~, 
}
define the affine conformal algebras $\widehat{\mathfrak{su}}(2)_{k-2}$ and $\widehat{\mathfrak{sl}}(2,\Real)_{k'+2}$, respectively, on the right--moving side. In addition, we have an  $(1,0)$--SCFT for the six--dimensional flat space--time part, with superfields  $(X^\gm, \gps^\gm_\mathsf{L})$, with $\gm=0,\ldots, 5$, regrouping the flat coordinate fields $X^\gm$ and their super partners $\gps^\gm_\mathsf{L}$. Furthermore, we have a system of 32  right--moving  Majorana--Weyl free fermions, $\gx_R^I$, $I=1,\ldots,16$, generating the target $SO(32)$ gauge and matter degrees of freedom.  Finally, we have to include a $(1,0)$--super--reparametrisation ghost system $(\varphi_\mathsf{L},\psi^\varphi_\mathsf{L})$.

%  classical asymmetric gauging

The worldsheet theory for the near--horizon limit of the warped Eguchi--Hanson heterotic background presented in section~\ref{sc:limit} was derived in~\cite{Carlevaro:2008qf}  as a WZW model for the asymmetric gauging:
\equ{ 
\Real^{1,5} \times 
\frac{SU(2)_k %/\Intr_2 
\times SL(2,\Real)_{k'} \times \Cplx^{16}}{U(1)_\mathsf{L} \times U(1)_\mathsf{R}}~:
\quad 
\big(g, g'; \gx^I_R \big) \ra
\big( 
\ee^{i \gs_3 \ga} g,  \ee^{i\gs_3 \ga} g' \ee^{i\gs_3 \gb}; \gx_R^I \ \ee^{-i \gb Q_I}
\big)~.
\label{cftback} 
}
The group action on the fermions is given in terms of a vector $\mathbf{Q}=(Q_1,\ldots,Q_{16})$, which is for notational simplicity denoted by the same symbol as the vector of magnetic charges  defining the line bundle~\eqref{gaugebund} in the heterotic supergravity discussion. 

% anomaly free asymmetric gauging

As the $U(1)$ factors are gauged in a chiral way, the resulting model will be anomalous, unless the following conditions, 
 \equ{ 
k = k' ~, 
\qquad 
k' + 2 =  \frac{\mathbf{Q}^2}{2}~, 
\label{CosetAnomalies}
}
are satisfied by the levels of the left-- and right--moving (super--)conformal algebras. The calculation of the anomalous terms for the gauged WZW model~\eqref{cftback} is similar to the derivation  in~\cite{Johnson:2004zq}.

% Identify the background 

Next we identify the geometrical background that this CFT describes. The asymmetric gauging \eqref{cftback} defines the coupling of the currents, 
\equ{\label{Rcurrent}
J_{\mathsf{L} } = J^3_{\mathsf{L} }  + K^3_{\mathsf{L} }~, 
\qquad 
J_{\mathsf{R} } = k_{\mathsf{R} } ^{3} + 
\frac{i}{2}\, \sum_{I=1}^{16} Q_I :\!\xi_{\mathsf{R} } ^{2I} \xi_{\mathsf{R} } ^{2I-1}\!:~, 
}
to the non--dynamical gauge fields $(A_\mathsf{L},\bA_\mathsf{L})$ and $(A_\mathsf{R}, \bA_\mathsf{R})$. Implementing this in the  $SU(2)_k \times \slr_{k'}$ action yields:
\equ{ \label{gdaction}
\begin{array}{rcl}
 S_\text{gauged}& =& {\ds  \frac{k-2}{8\pi} \int \d^2 z \,\big(\p\theta\, \pb \theta +  \p\psi\, \pb\psi +  \p\phi\, \pb\phi +2\cos\theta\, \p\phi\, \pb\psi\big) } 
 \\[2ex]
   &&  {\ds + \frac{k'+2}{8\pi} \int \d^2 z \,\big(\p\rho\, \pb \rho +  \p\psi'\, \pb\psi' +  \p\phi'\, \pb\phi' +2\cosh\rho\, \p\phi'\, \pb\psi'\big)  + S_{\text{Ferm}} + S(A)~.} 
\end{array}
}
Here $S_\text{Ferm}$ is the corresponding fermion action induced by the worldsheet supersymmetry, with simple derivatives, since we have grouped the torsion part of the covariant derivative for fermions in the gauge field contribution: 
\equ{
S(A) = \frac{1}{8\pi}  \int \di^2 z \,  \big[ 2i \bar A_{\mathsf{L}}   J_{\mathsf{L}}  +  2i J_{\mathsf{R}} A_{\mathsf{R}}
+(k-2) A_{\mathsf{L}}  \bar A_{\mathsf{L}}   - (k'+2) (A_{\mathsf{L}}  \bar A_{\mathsf{L}}  + A_{\mathsf{R}} \bar A_{\mathsf{R}} + 2\cosh\rho\, A_{\mathsf{R}} \bar A_{\mathsf{L}}  ) \big]~.
 \non}
When the anomalies are cancelled, one can simply integrate out the gauge fields over their algebraic equations of motion. After gauge fixing the $\slr$ fields $\psi'=\phi'=0$ and for large level $k$,~\footnote{This in order to compare with the background~(\ref{DSgeom})--\ref{eqDSL}), which is  an exact solution to the Bianchi identity~(\ref{bi}) only in the large brane charge limit, but receives corrections for finite brane charge values, in accordance with the supergravity discussion of Section~\ref{sc:general}.} we obtain the following action for the bosonic sector:
 \equ{\label{WSactionCFT}
 \begin{array}{rcl}
  S_{\text{bos.}} & = &{\ds  \frac{k}{8\pi}  \int \di^2 z \,  \big[ 
  \p\rho\, \pb \rho + 
  \p\theta\, \pb \theta + \sin^2\theta\, \p\phi\, \pb\phi +\tanh^2\rho\, (\p\phi +\cos\theta\, \p\phi  )\, (\pb\psi +\cos\theta\, \p\phi   ) }
  \\[2ex]
  && \hspace{9.5cm}+ \cos\theta\,  (\p\phi \pb\psi -\p\psi\, \pb \phi) \big] \,.
    \end{array}
 }
 This is precisely the worldsheet action one obtains when taking the background \eqref{DSgeom} and \eqref{eqDSL} in the double scaling limit \eqref{DSL} (normalized by a factor of $1/2\pi\ga'$), provided one identifies the five--brane charges: 
\equ{
\mathcal{Q}_5 = k~, 
\label{LevelBraneCharge} 
}
with the level $k$ of the WZW model, as can be established from the background solution~(\ref{DSgeom}). With this identification, the anomaly conditions~(\ref{CosetAnomalies}) yields the relation:
\equ{\label{kBI}
k = \frac{\mathbf{Q}^2}{2} - 2 \,,
}
which precisely reproduces the integrated Bianchi identity~(\ref{intBI}) for $c_2=1$, i.e.\ for the double--scaling limit of the generalized spin--connection $\gO_-$.  In particular, models with $\mathbf{Q}^2=4$ with vanishing five--brane charge and hence no--throat
opening in the geometry consistently  correspond to the flat $k=0$ case.

Moreover, from integrating the gauge fields we infer the existence of a non--constant dilaton:
 \equ{\label{dilCFT}
 \Phi(\rho) = \Phi_0 -\tfrac12 \ln \cosh\rho~.
 }
Finally, from the action for the fermions we extract the gauge field background: 
\equ{
\mathcal{A} =  \frac{\sigma_3^\mathsf{L} }{2\cosh\rho} \, Q_I \mathsf{H}^I~,
}
with the generators $\mathsf{H}^I$ defining the directions in the Cartan subalgebra $\mathfrak{h}(SO(32))$.

Notice at this point that we have not yet implemented in the WZW model the $\zi_2$ orbifold eliminating
the bolt singularity in the Eguchi--Hanson geometry~(\ref{metric}). This can best be understood from the blow down
limit of the theory, as we will see subsequently.

%%%%%%%%%%%%%%%%%%%%%%%%%%%%%%
\subsection{The WZW $\zi_2$ orbifold}
\label{sc:WarpedEHorbifold}

% connection to the R_Q geometry
The use of the radial coordinate $\cosh\rho =(r/a)^2$ in the near--horizon limit (see Section~\ref{sc:limit}), where the blow--up modulus $a$ is scaled away, does not allow to perform the blow down in  the WZW model by taking a continuous
limit. Starting from the background~(\ref{metric}) for $a=0$, the worldsheet theory can be constructed in this case from 
a linear $(1,0)$--super--dilaton theory together with an $\mathcal{N}=(1,0)$ $SU(2)_k$ WZW model.

\subsubsection*{Gauged fixed WZW model}

To understand the worldsheet origin of the $\zi_2$ orbifold, we go back to
the gauge fixing procedure we performed to obtain the action~(\ref{WSactionCFT}), where after integrating the gauge fields classically, use has been made of the $U(1)_L$ and $U(1)_R$ gaugings in\eqref{cftback} to choose the gauge 
\equ{\label{U(1)LRfixing} 
\psi'=\phi'=0~, 
\quad\text{i.e.}\quad 
g'_\text{fixed} = \ee^{\frac 12 \gs_1 \gr}~, 
\qquad 
\gr > 0~, 
}
for $g'$ generically given in \eqref{SUSLgroupElmts}. This gauge fixing does not fix the $U(1)_\mathsf{L}$ and $U(1)_\mathsf{R}$ transformations in \eqref{cftback} completely: Taking $\ga = \gb  = \pi\, s$, $s=0,1$ leaves $g'_\text{fixed}$ inert. This $\Intr_2$ group acts as $g \ra -g$ on the $SU(2)_k$ group elements in~(\ref{cftback}) and leaves the current algebra invariant. Since it results from imposing the reduced periodicity on the $S^3$ coordinate:  $\psi \sim \psi +2\pi$, to avoid a bolt singularity in the Eguchi--Hanson, as explained in 
Section~\ref{sc:leftinv}, this orbifold is in fact non--chiral. In order to preserve $(1,0)$ supersymmetry in six dimensions, we let the orbifold act on the right--moving sector of
the $SU(2)$ CFT, i.e.\ on representations of the bosonic $\widehat{\mathfrak{su}}(2)_{k-2}$ affine algebra.

Hence in this gauge, the CFT reproduces a $\Intr_2$ orbifold of the Callan--Harvey--Strominger (CHS) background~\cite{Callan:1991dj}  corresponding to a stack of $k$ heterotic five--branes on a $\ci^2/\zi_2$ singularity:
\equ{\label{ResidualZ2}
\Real^{5,1} \times \Real_\mathcal{Q} \times \frac {SU(2)_k \times \Cplx^{16}}{\Intr_2}~:
\quad 
\big(g, \gx_R^I\big)  \ra 
\big(-g, \gx_R^I \  \ee^{-\gp i\, Q_I}\big)~, 
}
where $\Real_\mathcal{Q}$ is the super--linear--dilaton theory $(\varrho, \gps^\varrho_\mathsf{L})$ with the non--compact direction $\varrho$  canonically normalized, i.e.\ with background 
charge $\mathcal{Q} = \sqrt{2/\ga' k}$, and the $(1,0)$--supersymmetric $SU(2)_k$ WZW model corresponds to the three--sphere of radius $\sqrt{\alpha'k}$ . In light of this it is not surprising that the two seemingly very different partition functions for (4.4) and (4.26) of Ref.\ \cite{Carlevaro:2008qf} are in fact the same: They are the partition functions corresponding to the same background once described as a gauge theory and once in the gauge fixed version of an $SU(2)_k\times \slr_k$ WZW model. As can be see in \eqref{ResidualZ2} on the right--moving fermion $\gx_\mathsf{R}^I$ the residual orbifold action acts via the shift embedding defined by $\mathbf{Q}$. 

This analysis sheds new light on the appearance of the  $\zi_2$ orbifold in the blow down limit: 
In Ref.~\cite{Carlevaro:2008qf} the presence of the orbifold in the CFT was justified by requiring the existence of a Liouville potential in the twisted sector of the discrete representation  spectrum.
 The discussion here gives an understanding of the $\Intr_2$ orbifold from the viewpoint of the continuous representations. 
 
% Labels of su(2) and sl(2) representations 
\subsubsection*{The partition function for continuous representations}

The full partition function of the CFT describing the warped Eguchi--Hanson background in the double scaling limit receives contributions from both discrete and continuous $\slr/U(1)$ representations: Continous representations of (bosonic) $\slr_{k'+2}/U(1)$ are labelled by a complex spin $J= \frac12 +ip$, given in terms of the continuous momentum $p\in \er_+$. They have eigenvalue $M\in \zi_{2k}$ under the current $k^3_{\mathsf{R}}$. In the warped Eguchi--Hanson, these states correspond to massive modes extending in the bulk of the space but which are still (delta--function) normalizable and are concentrated away from the resolved $\mathbb{P}^1$. In contrast,  discrete (bosonic) representations have half integral
spin values $\sfrac12 < J < \sfrac{k'+1}{2}$. Note that $J=\sfrac12 , \sfrac{k'+1}{2}$ correspond to boundary representations. Their $k^3$
eigenvalue is related to the spin through $M=J+r$, $r\in \zi$.  These representations correspond to states localized in the vicinity of the resolved  $\mathbb{P}^1$.

The partition function for continuous $\slr/U(1)$ representations can be written down for a generic line bundle of the type~(\ref{bs}), while
for discrete representations it has to be determined case by case. As shown in~\cite{Carlevaro:2008qf}, for a generic bundle vector $\mathbf{Q}$
the $SU(2) \times \slr$ coset WZW model~(\ref{WSactionCFT}) can be realized as CFT with enhanced $(4,0)$ worldsheet supersymmetry,
leading for the present implementation of the line bundle to the following partition function for continuous representations:
\begin{multline} \label{pfbd}
\dsp 
Z_{\text{cont.\,repr.}}(\tau) = 
\frac{1}{(4\pi^2 \alpha'\tau_2)^2} \frac{1}{(|\eta|^2)^4}\,  
\int_0^\infty  \di p \, 
\frac{\dsp \big(|q|^2\big)^{\frac{p^2}{k}}}{\dsp |\get|^2}\,
\frac12 \sum_{\gamma,\delta=0}^1\, \sum_{2j=0}^{k-2} 
\ee^{\pi i \delta \left( 2j +\frac{k-2}{2}\gamma\right) }
 \chi^j  \bar\chi^{j+\gamma \left( \frac{k-2}{2}-2j\right)} \, \times 
 \\[1ex] \dsp 
 \times\, \frac12 \sum_{a,b=0}^1 (-)^{a+b} 
 \frac{\dsp \vartheta\oao{a\,\mathbf{e}_4}{b\,\mathbf{e}_4}}{\dsp \eta^4}
 \, \frac12 \sum_{u,v=0}^1
 \ee^{-\frac{\pi i}{4} \mathbf{Q}^2 \gamma\delta} \, 
 \frac{\dsp \bar\vartheta \oao{u \,\mathbf{e}_{16} +\gamma \,\mathbf{Q}}{v\, \mathbf{e}_{16}+\delta\, \mathbf{Q}}}{\dsp \bget^{16}}~. 
\end{multline} 
Let us briefly explain to which field the various factors in this partition function belong to: The first factors proportional to $1/(\gt^2_2 (|\get|^2))^2$ correspond to the free CFT of the coordinate fields $X^\gm$ in light--cone--gauge. The integral of $(|q|^2)^{p^2/k}/|\get|^2$ is due to the linear dilation $\gr$, and can alternatively be expressed in terms of continuous $\slr/U(1)$ characters. The affine $\widehat{\mathfrak{su}}(2)_{k-2}$ characters $\chi^j(\tau)$ depend on the (half--)integral spin $0 \leq j \leq \sfrac{k-2}{2}$ and are explicitly given in~\eqref{su2-char} and \eqref{chi-lim}. Since the $\Intr_2$ orbifold neither acts on the currents nor on the left--moving fermions $\gps_\mathsf{L}^\gm, \gps_\mathsf{L}^\ga, \gps_\mathsf{L}^\gr$, the corresponding contribution in~(\ref{pfbd}) result in the free partition function, defined in terms of the genus--4 theta functions \eqref{genusnTheta}, with the sum over the spin--structures labeled by $a,b=0,1$. Finally, the last factor involving the genus--16 theta functions results from the right--moving fermions $\gx_\mathsf{R}^{I}$ all with the same spin--structure. 

In comparison to the partition function written in~\cite{Carlevaro:2008qf} here we use a bundle vector $\mathbf{Q}$ in~(\ref{pfbd}) which in the normalization of that paper would be written as  $\sfrac 12 \mathbf{Q}$. 
This requires  an extra $ \ee^{-\frac{\pi i}{4} \mathbf{Q}^2 \gamma\delta}$ phase in the second line of~(\ref{pfbd}) to ensure that the twisting by $\sfrac12 \mathbf{Q}$ of the $SO(32)_1$ partition function results in a
$\zi_2$ automorphism of the $SO(32)_1$ lattice and modular invariance 
(see e.g.~\cite{Kawai:1986ah,Senda:1987pf,Scrucca:2001ni,GrootNibbelink:2003zj}).

As argued before the partition function also represents the partition function for the warped $\ci^2/\zi_2$ geometry by virtue of the 
gauging procedure~(\ref{U(1)LRfixing}). Mathematically, this equivalence can be shown by using the identity~(\ref{idSU2}).
Physically, this blow down limit  corresponds to a stack of $k$ heterotic five--branes at $r=0$ in the coordinate system of~(\ref{metric}). Their back--reaction on the geometry opens an infinite throat, which microscopically does not allow for the presence of localized modes in the spectrum. It is also worthwhile noting that because of the of the presence of an infinite throat at $r=0$, the $\zi_2$ orbifold has no fixed point in the warped albeit singular geometry. This is mirrored by  the form the $\zi_2$ orbifold takes in the right--moving
$SU(2)_{k-2}$, where it acts as shift orbifold on the $SU(2)$ spins. 

\subsubsection*{Line bundle vector conditions}

To determine the massless spectra we are primarily interested in the discrete representations of the CFT corresponding to the warped Eguchi--Hanson space in the double scaling limit. The partition function for such discrete representations, completing~(\ref{pfbd}) into the full partition function for the resolved geometry can in principle be derived but turns out to be rather complicated. For the class of models with $\mathbf{Q}=(2,2q,0^{14})$, $q\in \en$, the explicit form as been determined in~\cite{Carlevaro:2008qf}. 

Nevertheless, the partition function \eqref{pfbd} for the continuous representations still proofs quite useful in order to derive some consistency conditions on the input parameters $k$ and $\mathbf{Q}$ that hold for discrete representations as well. In particular, in order for \eqref{pfbd} to encode the standard GSO$_R$ projection we need to require that 
\equ{\label{sumQ}   
\frac 14\, \mathbf{e}_{16}\cdot \mathbf{Q} = 
\frac 14 \sum_I Q_I = 0 \text{ mod } 1 \,. 
}
This condition is more restrictive than the condition on $c_1(\mathcal{L})$ in supergravity~\eqref{bs}. It is nevertheless compatible
with the consistency condition for gauge shifts in heterotic CFTs on $\ci^2/\zi_2$ found in~\cite{GrootNibbelink:2003zj}.

The other condition bears on the compatibility between the $\frac12 \mathbf{Q}$ twist and the $\zi_2$ orbifold, which, in particular, requires that the Ramond sector of~(\ref{pfbd}) is modular invariant under the transformation $\tau \rightarrow \tau +2$. This is ensured if the following condition is met~\cite{Vafa:1986wx,Ibanez:1987pj}:
\equ{\label{Q24}
  \frac 14 \mathbf{Q}^2 = 0 \text{ mod } 1 \,.
}
In particular, from this condition we deduce that consistency with the $\zi_2$ orbifold requires the level of the affine algebras, and subsequently the five--brane charge, to be even in the CFT:
\equ{
 \mathcal{Q}_5 = k =  0 \text{ mod } 2
 }
Finally,  by looking at the fermionic sector of the  partition function~(\ref{pfbd}), we can give a microscopic characterization of the distinction between gauge bundles with or without vector structure arising in supergravity because of the Dirac quantization condition~(\ref{dirac}):
\begin{itemize}
\item Models with $\mathbf{Q} \in \zi^{16}$ support a gauge bundle with vector structure: When $\mathbf{Q}$ has $m$ odd entries, the ground state  in the twisted NS--sector is equivalent to an R--groundstate for $m$ complex fermions. Otherwise, for even integral  $\mathbf{Q}$ the twist by $\mathbf{Q}$ factorizes in~(\ref{pfbd}), so that twisted and untwisted sector NS--grounds states are equivalent.
\item Models with $\mathbf{Q} \in \zi^{16} + \sfrac12\, \mathbf{e}_{16}$ support a gauge bundle without vector structure: On the CFT
side the twisted sector groundstate is described by operators $\text{exp}(\sfrac i2 \mathbf{S} \cdot \mathbf{X}_\mathsf{r} )$, with
$\mathbf{S} \in \zi^{16} + \sfrac12\, \mathbf{e}_{16}$, which are not spin fields.
\end{itemize}
In the following we will mainly concentrate on models with integral bundle vectors, due to their simpler groundstates.
\subsection{Marginal operators in the heterotic warped Eguchi--Hanson CFT}
\label{sc:MasslessCFTspec} 
% tensor product of left and right moving vertex operators

The spectrum of massless states of the CFT in the double--scaling limit of the warped Eguchi--Hanson compactification can most easily be computed by determining its set of marginal vertex operators. A target--space state in six dimensions, such as a hyper multiplet or gauge multiplet, is thus described in the CFT by a specific vertex operator $\mathcal{V}_\text{6D}$. This operator contains free worldsheet $(1,0)$--superfields $(X^\mu,\psi^\mu_\mathsf{L})$, $\mu=0,...,5$, and the reparameterization super--ghost system  $(\varphi,\psi_\mathsf{L}^\varphi)$, whose  SCFTs factorize. 
 The contribution from the internal CFT is packaged in a vertex operator we denote by  $\mathcal{V}$. It decomposes in the tensor product of a left--moving operator 
$\mathcal{V}_\mathsf{L}$, which encodes the contribution of the $(SU(2)_{k}/U(1) )\times( \slr_{k'}/U(1))$ SCFT and the right--moving  vertex operator $
\mathcal{V}_\mathsf{R}$ of the $SU(2)_{k-2} \times \slr_{k'+2}/U(1) \times SO(32)_1/U(1)$ CFT.

\subsubsection*{Vertex operators for massless 6D states}

Since we are looking for massless states in six dimensions, we require the momenta $p_\mu$ associated to the space--time target fields to be light--like: $p_\mu p^\mu=0$ and  $\mathcal{V}$ to be marginal, i.e.\ with left-- / right--conformal weights $(\Delta, \bar\Delta)=(1,1)$. The hyper / vector multiplets are then described in the CFT by the following vertex operators:
\equ{
\mathcal{V}_\text{6D} = \ee^{-\varphi}\,  \mathcal{V}\, 
\begin{cases}  {\ds \ee^{ip_\mu X^\mu}}~, & \text{hyper multiplet}~, 
\\[1ex] 
 \psi^\mu~, & \text{gauge multiplet}~, \end{cases}
\qquad 
\mathcal{V} = \mathcal{V}_\mathsf{L} \otimes \mathcal{V}_\mathsf{R}~, 
\quad 
\gD(\mathcal{V}_\mathsf{L}) = \gD(\mathcal{V}_\mathsf{R}) = 1~.
\label{Marginality} 
}
The marginality condition for $\mathcal{V}_\mathsf{R}$ can then be satisfied by tensoring (anti--)chiral primaries of $SU(2)_{k}/U(1)$ and $\slr_{k'}/U(1)$, and
for $\mathcal{V}_\mathsf{L}$ primary operators of  $SU(2)_{k-2}$ and $\slr_{k'+2}/U(1) \times SO(32)_1/U(1)$.  In the asymptotic limit $\varrho\rightarrow \infty$, the corresponding vertex operators acquire a particularly simple form, as the  $SO(32)_1$ part becomes a free field theory. We give the operators relevant
for the computation of the spectrum of massless states in Table~\ref{tb:ConfWeights}, along with their conformal weights.

\begin{table}[!t]
\[
\renewcommand{\arraystretch}{1.5} 
\arry{c}{
\arry{|c||c c|c c| c c |}{
\hline 
~\text{Left--moving}~  & \multicolumn{2}{|c|}{~~~\text{(anti--)chiral}~\frac{SU(2)_k}{U(1)_\mathsf{L}}~\text{primaries}~~~} 
& \multicolumn{2}{|c|}{~~~\text{(anti--)chiral}~\frac{SL(2,\Real)_{k'}}{U(1)_\mathsf{L}}~\text{primaries}~~~~} & \multicolumn{2}{|c|}{\Real^{1,5}~\text{fields}} 
\\
\text{fields} & C_{\mathsf{L}\, j} & A_{\mathsf{L}\,j} & C'_{\mathsf{L}\,J} & A'_{\mathsf{L}\, J} & \der X^\gm & \gps_{\mathsf{L}}^\gm 
\\\hline\hline
\gD & \frac12 -\frac{j+1}{k} &  \frac{j}{k} &  \frac J{k'} &  \frac12 -\frac{J-1}{k'}  & 1 & \frac 12 
\\\hline 
}
\\[-3.75ex] \\
\arry{|c||c|c|c|c|}{
\hline 
\text{Right--moving}  & {SU(2)_{k-2}}~\text{primaries} &  \multicolumn{2}{|c|}{\frac{\slr_{k'+2}}{U(1)_\mathsf{R}}~\text{primaries}   \times \frac{SO(32)_1}{U(1)_\mathsf{R}}~\text{torus fields}} 
&  \Real^{1,5}~\text{fields}
\\[0.2cm] 
\text{fields} & V_{\mathsf{R}\, j_\text{sh},m}  & \ee^{-\sqrt{\frac{2}{\alpha'k'}}J \varrho - i \mathbf{P}_\text{sh}\cdot \mathbf{X}_\mathsf{R}} & 
 \mathbf{Q} \cdot \bar\partial X_\mathsf{R}\,  \ee^{-\sqrt{\frac{2}{\alpha'k'}}J \varrho - i \mathbf{P}_\text{sh}\cdot \mathbf{X}_\mathsf{R}} &  \bder X^\mu
\\ \hline\hline
\bgD  & \frac{j_\text{sh}(j_\text{sh}+1)}{k} &  -\frac{J(J-1)}{k'}  + \frac 12\, \mathbf{P}_\text{sh}^2  & 1-\frac{J(J-1)}{k'}  + \frac 12\, \mathbf{P}_\text{sh}^2  & 1
\\ \hline 
}
}
\renewcommand{\arraystretch}{1} 
\]
\caption{In this table we give the primary operators of interest for discrete representations of the warped Eguchi--Hanson CFT in the asymptotic limit along with their conformal weights. The right--moving primaries correspond to $\slr_{k'+2}/U(1)$  representations $(J,M)$ with $M = J +r$, $r\in \en$, and $M = J-1$ respectively. The AdS$_3$ radial direction
$\varrho$ is canonically normalized. As a zero--mode it is non--chiral, but for simplicity
has been grouped together with the right--moving fields. The shifted  right--moving momentum $\mathbf{P}_\text{sh}$ and spin $j_\text{sh}$ are given in  \eqref{Psh} and \eqref{jsh}, respectively.}
\label{tb:ConfWeights}
\end{table}

% Disentanglement limit and bosonization

\subsubsection*{Vertex operators for discrete representations} 

The marginality condition for normalizable states in~(\ref{Marginality}) can only be satisfied by operators corresponding to discrete $\slr/U(1)$ representations, thus by states which are localized in the vicinity of  blown--up $\mathbb{P}^1$.
We now describe the properties of the vertex operators for discrete representations in detail. The $U(1)_\mathsf{R}$ which is gauged in~(\ref{cftback}) has direction in the Cartan subalgebra of  $SO(32)_1$ corresponding to the level $k+2$ current~(\ref{Rcurrent}):
\equ{\label{ucurr}
J_\mathsf{R} = k_\mathsf{R} ^3 +  \frac{i}{\sqrt{2\alpha' }} \,\mathbf{Q} \cdot  \bar\partial \mathbf{X}_\mathsf{R}  \,,
} 
where we have bosonized the right--moving fermions $\gx^I_\mathsf{R}$ via 
\equ{ \label{bosonization} 
 :\!\xi_\mathsf{R}^{2I-1}\xi_\mathsf{R}^{2I}\!: \, = 
 \sqrt{\tfrac{2}{\alpha'}} \, \bar\partial X^I_\mathsf{R}~,  \qquad I=1,\ldots,16 \,.
}
For a general  bundle vector $\mathbf{Q}$ the $\slr$ contribution to the partition function for discrete representations is given in terms of bosonic  $\slr_{k'+2}/U(1)$ characters. As mentioned before, in addition to their (half-)integral spin $J$, such discrete representations are further characterized by their eigenvalue  $M$ under $k^3$. Given expression~(\ref{ucurr}) we have:
\equ{ \label{Mcond}
M= \frac12 \mathbf{Q}\cdot \mathbf{P}_{\text{sh}} \,,  \qquad   \text{with} \qquad   M = J + r\,, \quad r\in \zi \,,
}
in terms of the sixteen dimensional charge vector $\mathbf{P}_{\text{sh}}$ of the gauge fermions~(\ref{Psh}).

By exploiting the non--compactness of the group $SL(2,\Real)$ it is possible to define a limit in which one can obtain rather explicit forms for these vertex operators~\cite{Aharony:2004xn}. For $SL(2,\Real)_{k'}/U(1)$ operators it is standard to use the non--compact radial direction $\varrho$
of AdS$_3$, with the canonical normalization $\varrho= \sqrt{\alpha' k'}\, \rho/2 $ with respect to the non--compact $\slr_{k'}$ coordinate in~(\ref{SUSLgroupElmts}).
Then, in the asymptotic limit $\varrho \ra \infty$ and using the gauge \eqref{U(1)LRfixing} the right--moving $SL(2,\Real)_{k'}/U(1)$ primary
operators assume a free  field expression, as given in the second line of Table~\ref{tb:ConfWeights}, where the bosonized vertex operators associated to the fermions  $\gx_\mathsf{R}^I$ are given in terms of so--called shifted momenta:
\equ{\label{Psh}
\mathbf{P}_\text{sh} = \mathbf{P} + \frac{\gamma}{2}\,  \mathbf{Q}~, 
\qquad 
\mathbf{P} \in \Lambda_{16} =
 \Big\{
\mathbf{N} + \frac{u}{2}\, \mathbf{e}_{16} ~\Big|~  
\mathbf{N} \in ~ \zi^{16} \, ;
u = 0,1 \Big\}~,  \quad \text{and} \quad  \gamma=0,1 \,,
}
These shifted momenta  encode NS ($u=0$) or R ($u=1$) boundary
conditions  of the corresponding string state and whether the state belongs  to the untwisted ($\gamma=0$) or twisted sector 
($\gamma=1$)  of the $\zi_2$ orbifold. In analogy to $\mathbf{P}_\text{sh}$ we denote the right--moving $SU(2)$ spin by:
\equ{\label{jsh} 
j_\text{sh} = j + \gg \Big( \frac {k-2}2 -2 j \Big)~, 
}
which takes into account the twisted and untwisted sector simultaneously.

% Consistent gauging

Not all marginal vertex operators $\mathcal{V}$ correspond to physical target space states. Since the vertex operator has to be inert under the asymmetric gauging~\eqref{cftback}, this enforces that the $SU(2)$ and $SL(2,\Real)$ levels have to be identified. This corresponds to the first anomaly condition~(\ref{CosetAnomalies}) for the coset CFT, i.e.\  
\equ{\label{MQ}
k'=k~.
}
In addition, the left-- and right--moving vertex operators, $\mathcal{V}_\mathsf{L}, \mathcal{V}_\mathsf{R}$, have to satisfy their respective GSO projections. Finally, the vertex operator $\mathcal{V}$ as a whole has to be orbifold invariant. 

%%%%%%%%%%%%SL weights

\subsubsection*{Right--moving $\boldsymbol{\slr/U(1)}$ representations and conformal weights}

In Table~\ref{tb:ConfWeights} the asymptotic limit 
$\varrho \rightarrow \infty$ gives rise to operators where  the dilaton and $SO(32)$ torus field dependences factorize.
However, not all such operators are in the CFT defined in~(\ref{cftback}). Only those that fall in representations
of the right--moving $\slr_{k'+2}/U(1) \times SO(32)_1/U(1)$ conformal algebra are. This in particular restricts the charge vector  $\mathbf{P}_\text{sh}$ to lie in the weight lattice of the unbroken gauge groups~(\ref{GaugeGroup}). 

To better understand the conformal weights of right--movers in Table~\ref{tb:ConfWeights}, let us recall that 
bosonic primaries of $\slr_{k'+2}/U(1)$ for discrete representation of spin $J$ and $k^3$ eigenvalue $M=J+r$, $r\in\zi$
have different expressions  depending on whether $r$ is positive or negative, namely:
\beq\label{discussr}
\begin{array}{lclcl}
r \geq 0 \;  : &  &{\ds  \bar\Delta = -\frac{J(J-1)}{k'} + \frac{M^2}{k'+2}}     \,; \\[10pt]
r < 0   \; : && \bar\Delta ={\ds  -\frac{J(J-1)}{k'} + \frac{M^2}{k'+2} -r}   \, .
\end{array}
\eeq
In particular,  $\slr_{k'+2}/U(1)$ primaries states with $r<0$  are affine descendants of bosonic primaries of lowest weight, i.e.\ with $J=M$, obtained
by applying the $\slr_{k'}$ generator  $k^-_{-1}= k^1_{-1} -i k^2_{-1}$ and thus have vacuum state $(k^-_{-1})^{-r}|J,J\rangle$ (see~\cite{Kiritsis:1986rv,Israel:2004xj, Eguchi:2004yi} for the $\mathcal{N}=2$ $\slr/U(1)$ coset and~\cite{Fotopoulos:2004ut} for the bosonic one).
These affine descendants correspond to vertex operators containing derivatives of target space fields. One should note in particular that taking similarly descendants of primaries with $M>J$ does not give a primary state.
 
Since right--moving fermions from the $SO(32)_1/U(1)$ coset theory have conformal weights given by:
\equ{
\bar\Delta 
%= \frac{1}{2}  \mathbf{P}_\text{sh} \cdot  
%\left( \mathbf{P}_\text{sh}  -  \mathbf{Q}\, \frac{ (\mathbf{Q} \cdot \mathbf{P}_\text{sh})}{ \mathbf{Q}^2}  \right) 
= \frac{1}{2}  \mathbf{P}_\text{sh}^T   
\left( \Id  -  \frac{ \mathbf{Q}\, \mathbf{Q}^T}{ \mathbf{Q}^2}  \right) 
\mathbf{P}_\text{sh} 
= \frac12 \mathbf{P}_\text{sh}^2 -  \frac{M^2}{k'+2}~, 
}
where the second equality is obtained by using the anomaly condition~(\ref{CosetAnomalies}) and expression~(\ref{Mcond}). 
Consequently, the total conformal weight of  a vertex operator composed of the product of a right--moving $\slr_{k'+2}/U(1)$ primary and 
a $SO(32)_1/U(1)$ state sums up to:
\equ{\label{cwtot}
\bar \Delta = \left\{ 
\begin{array}{ll}
{\ds -\frac{J(J-1)}{k'}  +  \frac12 \mathbf{P}_\text{sh}^2 }  \;,&   (M \geq J)  \\[0.4cm]
{\ds -\frac{J(J-1)}{k'}  + J-M +  \frac12 \mathbf{P}_\text{sh}^2 }  \;,  \quad&  (M < J)
\end{array}
\right.
}
The operators
$ \ee^{-\sqrt{\frac{2}{\alpha'k'}}J \varrho-i \mathbf{P}_\text{sh}\cdot \mathbf{X}_\mathsf{R}}$ correspond to primaries with
$r\geq 0$, while 
$\mathbf{Q} \cdot \bar\partial X_\mathsf{R}\,  \ee^{-\sqrt{\frac{2}{\alpha'k'}}J \varrho-i \mathbf{P}_\text{sh}\cdot \mathbf{X}_\mathsf{R}}$,
which contain only simple derivatives of target--space fields,  correspond to primary states $k^-_{-1}|J,J\rangle$ with $r=-1$ and thus with fixed $k^3$ eigenvalues $M=J-1$. 
Along this line we could in principle also consider bosonic $\slr_{k'+2}/U(1)$ primaries with $r < -1$. However, in this case the contribution $J-M$ to the conformal weight~(\ref{cwtot})  can be shown  to always lead to $\bar\Delta_\text{tot} > 1$ for the whole $SU(2)_k \times \slr_{k'}$ operator,
once the marginality condition $\Delta_\text{tot} = 1$ is satisfied for left--movers.

%%%%%%%%%%%%%%%%%%%%%%%%%%%%%%% 

\subsection{Hyper multiplets}
\label{sec:cftmodels}

In order to construct vertex operators corresponding to massless hypermultiplets in supergravity, we now construct the internal $\mathcal{V}$ operator in~(\ref{Marginality}) by  looking for marginal operators in the 
CFT obtained by tensoring superconfromal chiral or anti--chiral primaries of  $SU(2)_k/U(1) \times SL(2,\Real)_{k'}/U(1)$ with primaries of $SU(2)_{k-2} \times \slr_{k'+2}/U(1) \times SO(32)_1/U(1)$.

\subsubsection*{Left--moving vertex operator}  

The left--moving part of such operators can easily be realized by exploiting the super--conformal symmetry on the left. The idea is to start from a primary of $SU(2)_k/U(1) \times SL(2,\Real)_{k'}/U(1)$
with $\Delta =\frac12$ and subsequently take a descendant thereof by applying the supercharge 
$G_{-1/2} = G_{-1/2}^{su\,+}+ G_{-1/2}^{su\,-} +G_{-1/2}^{sl\,+}  + G_{-1/2}^{sl\,-} $ of the $(1,0)$  subalgebra of 
the total $(4,0)$  left-moving super-conformal algebra.

Table~\ref{tb:ConfWeights} provides us with two possibilities to combine operators of the supersymmetric $SU(2)_k/U(1)$ and $SL(2,\Real)_{k'}/U(1)$  to form the scalars of the hyper multiplet:  either one tensors two chiral primaries $C_{\mathsf{L}\, j} \otimes C'_{\mathsf{L}\, J}$, or two anti--chiral primaries  $A_{\mathsf{L}\,j} \otimes A'_{\mathsf{L}\, J}$. Note that
(anti-)chiral primaries of $SU(2)_k/U(1)$ have fixed $m=2(j+1)$ ($m=2j$) and odd (even) fermion number, while (anti-)chiral primaries of $SL(2,\Real)_{k'}/U(1)$   have bosonic charge $M=J$ ($M=J-1$) and  even (odd) fermion number. For simplicity, we deliberately omit the $m$, $M$ and fermion number labels in our notation for (anti-)chiral primaries. For more details on $SU(2)/U(1)$ and $SL(2,\Real)/U(1)$ characters and representations, see Appendix A in~\cite{Carlevaro:2012rz}, for instance.

Using the anomaly condition, $k=k'$, in~\eqref{MQ} and the conformal weights listed in Table~\ref{tb:ConfWeights}, we may obtain  $\Delta =\frac12$ by identifying $J=j+1$. Since the vertex operator thus constructed are complex fields,
we can combine them to obtain the two left--moving complex scalars expected from the lowest component of a hyper multiplet in 6 dimensions.
The left--moving part of the vertex operator~(\ref{Marginality}) thus reads:
\equ{
\mathcal{V}_\mathsf{L} = G_{-1/2} \big( C_{\mathsf{L}\, j} \otimes C'_{\mathsf{L}\, j+1} \oplus 
A_{\mathsf{L}\,j} \otimes A'_{\mathsf{L}\,j+1} )  \,.
}

\subsubsection*{Right--moving vertex operator} 

The right--moving part $\mathcal{V}_\mathsf{R}$ of the vertex operators are obtained by tensoring a primary $V_{\mathsf{R}\, j_\text{sh},m}$ of the bosonic right--moving $SU(2)_{k-2}$ with one of the primaries of the right--moving ${\slr_{k'+2}}/U(1) \times SO(32)_1/U(1)$ CFT listed in Table~\ref{tb:ConfWeights}.  This gives rise to two families of vertex operators:
\enums{
\item[{\large \textit{i)}}]  {\bf Type $\mathcal{V}_\mathsf{R}^{(1)}$ operators:}
\\[2ex] 
The first type of vertex operators reads in the asymptotic limit:
\equ{\label{V1} 
\mathcal{V}_\mathsf{R}^{(1)}  = 
\ee^{-\sqrt{\frac{2}{\alpha'k'}} J \varrho -i \mathbf{P}_\text{sh}\cdot \mathbf{X}_\mathsf{R}} \otimes
V_{\mathsf{R}\, j_\text{sh};m}~,
}
with ${\slr_{k'+2}}/U(1)$ charge \eqref{Mcond} with $ r\in  \en$. 
For the reason mentioned above only representations with $r\geq 0$ may lead to massless states. The marginality condition
\begin{equation}
\bar\Delta  =  \frac{j_\text{sh}(j_\text{sh}+1)}k - \frac{J(J-1)}{k'} + 
\frac{1}{2}\mathbf{P}_\text{sh}^2 = 1
\end{equation}
simplifies, by using condition \eqref{CosetAnomalies}, the definition of $j_\text{sh}$~(\ref{jsh}), together with the marginality condition on left--movers $J=j+1$:
\equ{ \label{marg1}
\frac{1}{2}\mathbf{P}_\text{sh}^2 +\Big(\frac{k-2}{4}-j\Big)\gg = 1~. 
}

\item[{\large \textit{ii)}}]  {\bf Type $\mathcal{V}_\mathsf{R}^{(2)}$ operators:}
\\[2ex] 
The second type of vertex operators we can construct from Table~\ref{tb:ConfWeights} are:
\equ{\label{V2}
\mathcal{V}^{(2)}_\mathsf{R}  = \mathbf{Q}\cdot \bar{\partial} \mathbf{X}_\mathsf{R}\,
\ee^{-\sqrt{\frac{2}{\alpha'k'}} J \varrho -i \mathbf{P}_\text{sh}\cdot \mathbf{X}_\mathsf{R}}
 \otimes
V_{\mathsf{R}\, j_\text{sh},m}\,.
}
 Their ${\slr_{k'+2}}/U(1)$ charge is fixed by the spin:
\equ{\label{spinM2}
M = \frac12 \mathbf{Q}\cdot \mathbf{P}_\text{sh} = J-1 \,,
   }
since their vacuum state is $k_{-1}^-| J,J \rangle$.  Finally, we readily obtain from Table~\ref{tb:ConfWeights} and the same algebra as before the marginality condition in this case:
\begin{equation} \label{marg2}
\bar\Delta  = 1+ \frac{1}{2}\mathbf{P}_\text{sh}^2 +\Big(\frac{k-2}{4}-j\Big)\gg  = 1~.  
\end{equation}
}

\begin{table}[t!]
\[
\renewcommand{\arraystretch}{1.5} 
\arry{|c||c|c|c|c|}{ \hline
& \multirow{2}{*} {$\text{mass--shell condition}$ } & \slr_{k+2}/U(1)_\mathsf{R}\text{ rep} &  \multirow{2}{*} {$\text{GSO}_\mathsf{R}\text{ projection}$} & 
 \multirow{2}{*} {$SU(2)_{k-2\,\mathsf{R}}  \text{ degeneracy}$} \\
& &   \text{and orbifold projection} & & \\ \hline \hline
&& & \multirow{4}{*} {$\mathbf{N}\cdot \mathbf{e}_{16}= 0 \text{ mod } 2$}  &   \multirow{4}{*} {$-j_\text{sh} \leq m \leq  j_\text{sh}$}  \\[-0.4cm]
\mathcal{V}^{(1)}_\mathsf{R}  &   {\ds \mathbf{P}_\text{sh}^2 +\Big(\frac{k-2}{2}-2j\Big)\gg = 2} & {\ds  M= \frac{ \mathbf{Q}\cdot \mathbf{P}_\text{sh}}{2} = j + \en^*} &  
 &  \\  
 &&&& \\[-0.4cm] \cline{1-3}
 &&&& \\[-0.4cm]
 \mathcal{V}^{(2)}_\mathsf{R}   &  {\ds  \mathbf{P}_\text{sh}^2 +\Big(\frac{k-2}{2}-2j\Big)\gg = 0 }& {\ds M= \frac{ \mathbf{Q}\cdot \mathbf{P}_\text{sh}}{2} = j }& & \\[-0.4cm]
 &&&&
 \\ \hline
 }
\renewcommand{\arraystretch}{1} 
\]
%\vspace{-0.2cm}
\caption{This table gives a compact summary of the conditions satisfied by the full massless CFT  spectrum for the warped Eguchi--Hanson geometry in the double scaling limit.
\label{tb:SpecSum}}
\end{table} 

\subsubsection*{GSO and orbifold projections}

In addition to the conditions discussed above the massless states undergo GSO$_\mathsf{R}$ and orbifold projections. Although we are considering discrete $\slr_{k'}/U(1)$ representations, 
they can be determined from the partition function~(\ref{pfbd}), as these projections are the same for continuous and discrete representations

% GSO_R projection
Counting the fermionic number of~(\ref{V1}) and~(\ref{V2}) gives us the following  GSO$_\mathsf{R}$ projection:
\equ{\label{GSOR} 
\frac 12 \mathbf{e}_{16}\cdot \mathbf{P}_\text{sh}  = 0 \text{ mod } 1 \quad  \Rightarrow  \quad
\frac 12 \mathbf{e}_{16}\cdot \mathbf{N}
= 0 \text{ mod } 1~, 
}
where we have in particular used~(\ref{bosonization}) and~(\ref{sumQ}). For $\cV_\mathsf{R}^{(1)}$ the same GSO$_\mathsf{R}$ projection can alternatively be retrieved by extracting the $v$ dependent phases from the partition function~\eqref{pfbd}.

% Orbifold projection
Finally, we have to ensure that the states constructed here are invariant under the $\Intr_2$ orbifold action \eqref{ResidualZ2} which acts as%
\equ{
X_\mathsf{R}^I \ra X_R^I + \pi\, Q_I~, \quad I=1,..,16 \,,
\qquad 
V_{\mathsf{R}\, j_\text{sh},m} \ra (-)^{2j + \frac {k-2}2 \gg } \, V_{\mathsf{R}\, j_\text{sh},m}~, 
}
on the constituents of the vertex operators.  Hence for the vertex operators \eqref{V1} and \eqref{V2} to be invariant, we require that 
\equ{\label{orbhyp}
\frac12 \mathbf{Q}\cdot \mathbf{P}_\text{sh} + \frac 18 \, \gg\, \mathbf{Q}^2 = j + \gg\, \frac{k-2}{4} \text{ mod } 1 \,.
}
The contribution $\frac 18 \, \gg\, \mathbf{Q}^2$ results from the so--called vacuum phase in the twisted sector. As it is universal, it can be read off from the partition function~\eqref{pfbd} for the continuous representations by identifying its $\delta$--dependent phases. Notice that
upon using the anomaly condition~(\ref{kBI}) and the definition of the charge $M$~(\ref{MQ}), condition~(\ref{orbhyp}) simplifies to:
\equ{
M = j \text{ mod } 1 \,.
}
Given the identification of left--moving spins $J=j+1$, this condition is trivially satisfied by right--moving $\slr_{k'+2}/U(1)$ discrete representations, which have $M=J+r$, $r\in\zi$.

%%%%%%%%%%%%%%%Hyper multiplet spectrum

\subsubsection*{The hyper multiplet spectrum}

By solving the algebraic system in Table~\ref{tb:SpecSum} subject to constraints~(\ref{sumQ}) and~(\ref{Q24}),  we are now in the position to compute in general the hyper multiplet spectrum for the heterotic warped 
Eguchi--Hanson CFT. We restrict our analysis to integer valued  bundle vectors defined as in~(\ref{shiftAnsatz}) (leaving the half--integer case for future analysis). These vectors are subject to the GSO$_\mathsf{R}$ and $\zi_2$ orbifold conditions,~(\ref{sumQ}) and~(\ref{Q24}), the first one corresponding to a more stronger version of the K--theory condition~(\ref{bs}) in supergravity, guaranteeing stability of the Abelian gauge bundle. For the given choice of models, these conditions translate to:
\equ{\label{cond3}
  \sum_{i=1}^n N_i\, p_i =  0 \text{ mod } 4 \,, \qquad \text{and} \qquad   \sum_{i=1}^n N_i\, p_i^2 = 0 \text{ mod } 4  \,.
}
Also, the level  of the affine algebras and the range of $SU(2)$ and discrete $\slr$ left spins are in this case given by:
\equ{ \label{p2_sim_k} 
 \mathcal{Q}_5 \equiv k =  \frac12   \sum_{i=1}^n N_i \, p_i^2 -2  \,,
\qquad    J-1= j = 0, \, \frac12 ,\, 1 , \, \frac32, \, 2 ,  \, \ldots \,, \frac14  \sum_{i=1}^n N_i \, p_i^2 -2   \,.
}
\noindent 
The hyper multiplet spectra are determined by the following vertex operators: 
%%%%%%%%%%%%V1
\noindent
\items{
\item[{\large \textit{i)}}]  {\bf Type $\mathcal{V}_\mathsf{R}^{(1)}$ operators:}
\\[2ex] 
We first consider vertex operators built on the right--moving $\mathcal{V}_\mathsf{R}^{(1)}$~(\ref{V1}). The 
marginality condition~(\ref{marg1}) for these operators leads to the mass--shell equation:
\equ{\label{Pms}
\mathbf{P}_\text{sh}^2 =2+ \Big[ 2(j+1)-\frac14  \sum_{i=1}^n N_i \, p_i^2 \Big]\gamma \,.
}
\\[2ex] 
\textbf{Untwisted sector $\boldsymbol{(\gamma=0)}$: } 
\\[1ex] 
All solutions to equation~(\ref{Pms}) which are GSO$_\textsc{R}$ and orbifold invariant  are in the untwisted NS sector and therefore characterized by a momentum $\mathbf{P}_\text{sh} =\mathbf{N} \in \zi^{16}$. Since in the untwisted sector we have $j=j_\text{sh}$, the multiplicity operators for such massless states simply counts the internal $-j \leq m \leq j$ degeneracy within given right--moving $SU(2)_{k-2}$ representation of spin $j$. Given the marginality condition $j=J-1$ for left--movers, the multiplicity sums over degeneracies for all spins $j$ satisfying relation~(\ref{Mcond}):
\equ{ \label{ind1}
 n_\mathbf{Q}(\mathbf{P}_\text{sh}) = \left\{
 \begin{array}{ll}
{\ds   \sum_{j=0}^{M-1} (2j+1) = M^2 = \frac{(\mathbf{Q}\cdot \mathbf{P})^2}{4} \,, } & \quad  {\ds  M \in \en^* \,,} \\[14pt]
  {\ds  \sum_{j-\frac12=0}^{M-\frac32} (2j+1) = M^2-\frac14 = \frac{(\mathbf{Q}\cdot \mathbf{P})^2-1}{4} \,, }& \quad {\ds  M \in \frac12+ \en \,,}
  \end{array}
  \right.
}
\\[20ex] 
\textbf{Twisted sector $\boldsymbol{(\gamma=1)}$:}
\\[1ex] 
All massless states are in the twisted NS  sector, in accordance with what we found for the untwisted spectrum. In particular, if $\mathbf{Q}$ has $m$ entries $p_i$ which are odd, these operators correspond to untwisted  \textsc{r} ground states for $m$ of the complex fermions.
In this case however, since the relation between the shifted momentum and the spin $j$ is fixed by the mass--shell condition, all $SU(2)$ spins $j$ which are solution to equation~(\ref{cond3}) are in one--to--one correspondence with  representations of the unbroken gauge group, determined by $ \mathbf{P}_\text{sh}$. In the twisted sector the multiplicity therefore takes the form:
\begin{equation} \label{ind2}
n_\mathbf{Q}(\mathbf{P}_\text{sh}) =  k-1-2j  =   \frac{k+4}{2} - \mathbf{P}_\text{sh}^2 \,.
\end{equation} 
In particular, the twisted singlet in Table~\ref{tab:twsec} corresponds to the Liouville operator giving the CFT description 
of the blow--up mode in the Eguchi--Hanson.

As was shown in~\cite{Carlevaro:2008qf}, requiring the presence of the Liouville operator in the spectrum
accounts for the necessity of having a $\zi_2$ orbifold in the coset CFT~(\ref{cftback}) for the resolved space, since this operator
is in the twisted sector generated by the orbifold. As a byproduct, the GSO$_\mathsf{R}$ invariance condition for the
Liouville operator precisely gives the gauge bundle stability condition~(\ref{sumQ}). The marginal deformation of the CFT generated by this operator 
is called the Liouville potential, and encodes non-perturbative worldsheet instanton effects.  Thus,  from the perspective of discrete representations, both the presence the $\zi_2$ orbifold and the gauge bundle stability condition depend on the existence of
the Liouville potential and hence of a non-perturbative (in $\alpha'$) completion of the theory.

\item[{\large \textit{ii)}}]  {\bf Type $\mathcal{V}_\mathsf{R}^{(2)}$ operators:}
\\[2ex] 
The marginality condition for these operators reads:
\equ{
\mathbf{P}_\text{sh}^2 = \Big[ 2(j+1)-\frac14  \sum_{i=1}^n N_i p_i^2 \Big]\gamma \,.
}
Requiring the right--moving $\slr_{k'+2}/U(1)$ states to be primaries fixes  $M = J-1= j$.  
\\[2ex] 
\textbf{Untwisted sector $\boldsymbol{(\gamma=0)}$:} 
\\[1ex] 
In the untwisted sector, the only GSO$_\mathsf{R}$ and orbifold invariant massless state is the gauge group singlet with zero shift momentum and therefore $n_\mathbf{Q}=1$. It corresponds to the asymmetric current--current operator which gives rise to the dynamical deformation resolving the singular background geometry in~(\ref{ResidualZ2}).  As such, it gives the CFT description of the volume modulus of the blown--up $\mathbb{P}^1$ and is always turned on together with the Liouville operator in Table~\ref{tab:twsec}. This state is given in the last line of Table~\ref{tab:untwsec}.
\\[2ex] 
\textbf{Twisted sector $\boldsymbol{(\gamma=1)}$: }
\\[1ex] 
In the twisted sector, only $SO(2N)$ singlet representations appear with multiplicities given by: 
\equ{\label{ind3}
n_\mathbf{Q}(\mathbf{P}_\text{sh}) =  \frac{k}{2} - \mathbf{P}_\text{sh}^2 \,.
}
In this case, there are no massless twisted states in the $2N$ of $SO(2N)$.
The fundamental of $SO(2N)$ can be fermionized to give a state
 $\xi^{a}_{\mathsf{R}\,-1/2}|0\rangle_\textsc{ns}$, where $\xi^{a}_{\mathsf{R}\,n+1/2}$ are oscillators from the free gauge sector 
$SO(2N)$. Combining this with the $\mathbf{Q}\cdot \bar{\partial} \mathbf{X}_\mathsf{R}$ always leads to massive states in any $U(N_i)$ representation.
}
%

%%%%%%%%%%%%%%%%%%%%%%%%%%%%%%
% Page with just tables 

 \newpage %\thispagestyle{empty} 

\begin{table}[ht!]
\footnotesize
\[
\renewcommand{\arraystretch}{1.4} 
\hspace{-0.25cm}
\begin{array}{|c|c|c|c|l|} \hline 
\multicolumn{5}{|c|}{\text{Gauge group:}~  U(N_1) \times \ldots U(N_n)  \times SO(2N)} \\  
\text{Representation} & \mathbf{P}_\text{sh}= \mathbf{P} =    &  \multicolumn{1}{|c|}{M} &  J=j+1 & \multicolumn{1}{|c|}{ n_{\mathbf{Q}}(\mathbf{P}_\text{sh})}  \\[4pt] \hline\hline
 &&&& 
 \\[-14pt]
 (...,\rep{1}, [\rep{N_i}]_2,\rep{1} ,...  ;\rep{1} ) &(...,0,\underline{1^2,0^{N_i-2}},0,...; 0^{N})  & p_i   &  1,2,..., p_i &  p_i^2   
 \\[8pt]   
\multirow{3}{*}{ $(...,\rep{1}, \rep{N_i},\rep{1},...,\rep{1},\rep{N_j},\rep{1},...; \rep{1} ) $} 
&\multirow{3}{*}{ $ (...,0,\underline{1,0^{N_i-1}},0,...,0, \underline{1,0^{N_j-1}},0,... ; 0^{N}) $} 
& \multirow{3}{*}{ $ {\ds \frac{p_i+p_j}{2}} $}  &  1,2,..., {\ds \frac{p_i+p_j}{2}}  &  {\ds \frac{(p_i+p_j)^2}{4}  }   \\[8pt]  

&&&{\ds  \frac32, \frac52,..., \frac{p_i+p_j}{2} }   &{\ds \frac{(p_i+p_j)^2}{4} -\frac14}    
\\[8pt]   
\multirow{3}{*}{ $(...,\rep{1}, \rep{N_i},\rep{1},...,\rep{1},\crep{N_j},\rep{1},...;\rep{1} ) $}   
& \multirow{3}{*}{ $(...,0,\underline{1,0^{N_i-1}}, 0,...,0,\underline{-1,0^{N_j-1}},0,...; 0^{N}) $} 
& \multirow{3}{*}{ $ {\ds \frac{p_i-p_j}{2}}$} &  1,2,..., {\ds \frac{p_i-p_j}{2} } & {\ds \frac{(p_1-p_2)^2}{4}}  
\\[8pt]    
 && &{\ds  \frac32, \frac52,..., \frac{p_i-p_j}{2}} &  {\ds \frac{(p_i-p_j)^2}{4}-\frac14}   
 \\[8pt]  
\multirow{3}{*}{ $(...,\rep{1}, \rep{N_i}, \rep{1},... ;\rep{2N} )$}
& \multirow{3}{*}{ $  (...,0,\underline{1,0^{N_i-1}},0,...;  \underline{\pm 1, 0^{N-1}})$} 
& \multirow{3}{*}{ ${\ds \frac{p_i}{2}} $}
&  1,2,...,  {\ds \frac{p_i}{2}}  & {\ds \frac{p_i^2}{4}}   
\\[8pt]  
&& &  {\ds  \frac32, \frac52,..., \frac{p_i}{2}}  & {\ds \frac{p_i^2}{4}-\frac14}     \\[6pt] \hline\hline
(\rep{1},...,\rep{1};\rep{1}) & (0^{16-N}; 0^N) &  0 &  1 & 1 \\[2pt] \hline
\end{array} 
\vspace{-2ex} 
\]
\caption{This table gives the spectrum of untwisted massless bosonic states of the heterotic warped Eguchi--Hanson CFT. This table covers all models for integer valued  bundle vectors $\mathbf{Q}$ given in~(\ref{shiftAnsatz}). The  third column distinguishes between spins $J$ corresponding to even or odd values of $M$.
The second horizontal double line separates states corresponding to the untwisted operators of type $\mathcal{V}_\mathsf{R}^{(1)}$ from the singlet state described by the the untwisted  operator $\mathcal{V}_\mathsf{R}^{(2)}$. 
%The latter corresponds to the asymmetric current--current operator and gives the CFT description of the volume of the blown--up two--cycle.
\vspace{-4ex} 
}
\label{tab:untwsec}
\end{table}
%%%%%%%%%%%%%%%%%%%%
\begin{table}[hb!]
\footnotesize
\[
\renewcommand{\arraystretch}{1.3} 
\hspace{-0.5cm}
\begin{array}{|c||c|c|c|c|l|} \hline 
 & \multicolumn{5}{|c|}{\text{Gauge group:}~  U(N_1) \times \ldots U(N_n)  \times SO(2N)}  \\  
& \text{Representation} & \mathbf{P}_\text{sh} = \mathbf{P} + \sfrac 12\, \mathbf{Q}~,~ \mathbf{P}=  &  J=j+1 & M &  \multicolumn{1}{|c|}{\mathbf{n}_{\mathbf{Q}}(\mathbf{P}_\text{sh})}    \\[6pt] \hline\hline
 &&&& &  \\[-14pt]
\mathcal{V}_\mathsf{R}^{(1)}  & \multirow{3}{*}{ $ (...,\rep{1}, [\rep{N_i}]_2,\rep{1} ,...  ;\rep{1} )  $}
& \multirow{3}{*}{ $(...,0,\underline{-1^2,0^{N_i-2}},0,...; 0^{N}) $}
  &  {\ds \frac k2- p_i +1 }& J  &  2p_i-1     \\[8pt]  
\mathcal{V}_\mathsf{R}^{(2)}  & & &   {\ds \frac k2-p_i  +2 } & J-1 &  2p_i-3   \\[8pt]  \hline
 &&&& &  \\[-12pt]
  \mathcal{V}_\mathsf{R}^{(1)}  & \multirow{3}{*}{ $(...,\rep{1}, \rep{N_i},\rep{1},...,\rep{1},\rep{N_j},\rep{1},... ; \rep{1} ) $} 
&\multirow{3}{*}{ $ (...,0,\underline{-1,0^{N_i-1}},0,...,0, \underline{-1,0^{N_j-1}},0,... ; 0^{N}) $}
& {\ds \frac{k-p_i-p_j}{2}+1 }  & J &  p_i+p_j-1   \\[8pt]
 \mathcal{V}_\mathsf{R}^{(2)}& &  & {\ds \frac{k-p_i-p_j}{2} +2 }  & J-1 & p_i+p_j-3  \\[8pt ] \hline
 &&&& &  \\[-12pt]
  \mathcal{V}_\mathsf{R}^{(1)}   & \multirow{3}{*}{ $(...,\rep{1}, \rep{N_i},\rep{1},...,\rep{1},\crep{N_j},\rep{1},...;\rep{1} ) $}   
& \multirow{3}{*}{ $(...,0,\underline{-1,0^{N_i-1}}, 0,...,0,\underline{1,0^{N_j-1}},0,...; 0^{N})   $}
   & {\ds \frac{k-p_i+p_j}{2}  +1} &  J  & p_i-p_j-1   \\[8pt]  
\mathcal{V}_\mathsf{R}^{(2)}&     &  & {\ds \frac{k-p_1+p_2}{2} +2 }  & J-1 & p_i-p_j-3 \\[8pt] \hline
 &&&& &  \\[-12pt]
 \mathcal{V}_\mathsf{R}^{(1)}    & (...,\rep{1}, \rep{N_i}, \rep{1},...;\rep{2N} ) &
 (...,0,\underline{-1,0^{N_i-1}},0,...;  \underline{\pm 1, 0^{N-1}})  
 &    {\ds \frac{k-p_i}{2} +1 } & J & p_i -1    \\[8pt] \hline
 &&&& &  \\[-12pt]
\mathcal{V}_\mathsf{R}^{(1)}  & ( \rep{1},..., \rep{1} ;\rep{1} )& 
(0^{16-N}; 0^N)
& {\ds \frac{k}{2} }  & J+1  & 1  \\[6pt] \hline
\end{array}
\]
\vspace{-2ex} 
\caption{This table gives the spectrum of twisted massless bosonic states of the heterotic warped Eguchi--Hanson CFT. It covers all models for integer valued  bundle vectors $\mathbf{Q}$ of the form given in~(\ref{shiftAnsatz}).}
\label{tab:twsec}
\end{table}

\newpage

%%%%%%%%%%%%%%%%%%%%%%%%%%

The full  massless spectra  for untwisted and twisted $\mathcal{V}_\mathsf{R}^{(1)}$ and $\mathcal{V}_\mathsf{R}^{(2)}$ operators are given in 
Tables~\ref{tab:untwsec} and~\ref{tab:twsec}.
In Table~\ref{tab:untwsec} we observe that for a marginal operator with half integral spin values, the overall multiplicity of massless states $n_\mathbf{Q}$ for a particular representation of the unbroken gauge group is reduced by a factor $\sfrac 14$ with respect to the integral spin case. 
This phenomenon is related to the absence of  $\slr/U(1)$ discrete boundary representations in the spectrum. These discrete representations appear as a branching of the continuous representations in the limit where their momentum $p \rightarrow 0^+$ and have $\slr/U(1)$ spins $J=\{ \sfrac12, \sfrac{1}{2} (k+1)\}$ (see for instance~\cite{Israel:2004xj}). However, one can check that these representations are projected out of the partition function~(\ref{pfbd}), as they would lead to states with an $SU(2)$ spin
outside the allowed range $0 \leq j \leq  \sfrac{1}{2} (k-2)$.

%%%%%%%%%%%%%%%%%Gauge multiplets
\subsection{Gauge multiplets}
\label{sec:gaugemu}

%%%%%%%%%%%%%%
\subsubsection*{Massless gauge states}

Generically, the vertex operator corresponding to a gauge field in space--time contains as $\er^{5,1}$ component the operator $\gps_\mathsf{L}^\gm$, $\mu=0,..,5$, in the NS sector, i.e.\ $\gps_\mathsf{L\,-1/2}^\gm|0\rangle_\textsc{ns}$. Since this operator already has conformal weight $\gD = 1/2$, the marginality condition for the left--moving vertex operator $\mathcal{V}_\mathsf{L}$ dictates we tensor $\gps_\mathsf{L}^\gm$ with a descendant of a   left--moving state with $j=0=J$, which corresponds to the tensor product of the identity operators in $SU(2)_k/U(1) \times \slr_{k'}/U(1)$. For $SU(2)_k/U(1)$ the identity representation is the anti--chiral primary with $j=0$. For $\slr_{k'}/U(1)$ in contrast,
the identity is non-normalizable, as its spin $J=0$ is outside of the range $\frac12 < J < (k+1)/2$ characterizing discrete representations. Then the identification of $\slr_{k'}$ spins entails on the right--moving side the identity representation
for the bosonic $\slr_{k'+2}/U(1)$, with $(J,M)=(0,0)$.

With this in mind, one can verify that the marginality condition for right--movers can only be solved in the untwisted NS sector, leaving again no  other choice but the identity representation for  $SU(2)_{k+2}$ with $(j,m)=(0,0)$. The complete vertex operators corresponding to massless gauge multiplets take the form:
\equ{\label{vgauge}
 \mathcal{V}_\text{g} =  \gps_\mathsf{L}^{\gm} \big(G_{-1/2}\, A_{\mathsf{L}\,0,0} \, \text{Id}'_{\mathsf{L}} \big) \otimes  \, e^{-i \mathbf{P}_\text{sh} \cdot \mathbf{X}_\mathsf{R}} 
 V_{\mathsf{R}\,0,0}~\,.
}
with the untwisted momenta $\mathbf{P}_\text{sh}$ satisfying the mass--shell condition:
\equ{ \label{gms}
\bar\Delta  = \frac12   \mathbf{P}_\text{sh}^2 = 1~, 
}
Since the $\slr_{k'+2}/U(1)$ charge has value $M=0$, the only solutions to~(\ref{gms}) are untwisted NS ground states carrying the adjoint representations of the massless unbroken gauge group, as can be seen from Table~\ref{tab:gauge}. In particular, by fermionizing the $\mathbf{P}_\text{sh}$--dependent constituent of the operator~(\ref{vgauge}) for solutions in Table~\ref{tab:gauge}, we observe that in the CFT
description the string states associated to massless gauge bosons are consistently constructed by tensoring the identity representations in  $SU(2)_{k-2} \times (\slr_{k'+2}/U(1))$ with states $\xi^I _{\mathsf{R}\, -1/2}\xi^J _{\mathsf{R}\, -1/2} |0\rangle_\textsc{ns}$,
which is what we expect.

As mentioned, the identity representation corresponds to a non--normalizable string state. As a consequence the wave--functions of the unbroken gauge group bosons do not have support on the resolved two--cycle of the warped Eguchi--Hanson space and thus correspond to a \textit{global} symmetry of the interacting theory localized on the blown--up $\mathbb{P}^1$. In Table~\ref{tab:gauge} we give the relevant representations for gauge multiplets as determined from the CFT for models defined by the line bundle vector~(\ref{shiftAnsatz}).

\begin{table}[t!]
\[
\renewcommand{\arraystretch}{1.5} 
\begin{array}{|c|c|c|c|}\hline 
\multicolumn{4}{|c|}{\text{Gauge group:}~ U(N_1)\times \ldots \times U(N_n) \times SO(2N)} \\ 
\text{Representation} &  \mathbf{P}_\text{sh}  &  M = \frac12 \mathbf{Q}\cdot \mathbf{P}_\text{sh}   & j = J  
\\ \hline\hline 
\multicolumn{4}{|l|}{\text{Massless non--normalizable gauge multiplets}} \\ \hline 
(\rep{1},\ldots ,\rep{1}, \rep{Ad}_{U(N_i)}, \rep{1}, \ldots ,\rep{1}; \rep{1})  & (0,\ldots, 0, \underline{1,\sm1,0^{N_i-2}}, 0 ,\dots,0 ;0^N)  &  0 & 0  \\
(\rep{1},\ldots ,\rep{1};\rep{Ad}_{SO(2N)})  & (0, \ldots, 0; \underline{\pm1^2,0^{N-2}}) &   0 & 0  \\[1ex] \hline
\multicolumn{4}{|l|}{ \text{Massive Abelian gauge field, with mass } \frac{4}{\sqrt{\alpha'(\mathbf{Q}^2 -4)}} } \\[1ex] \hline 
U(1)_\mathbf{Q} & (0^{16-N}; 0^N \ldots  0) &  0  & 1  \\ \hline
\end{array}
\]
\caption{This table give the non--normalizable massless and the massive $U(1)_\mathbf{Q}$ gauge multiplets.}
\label{tab:gauge}
\end{table}

% GS mass for a U(1) 
\subsubsection*{The massive $\boldsymbol{U(1)_\mathbf{Q}}$ gauge field}

As we recalled in Subsection~\ref{sc:spectra}, turning on a line bundle gauge background in six--dimensional heterotic compactifications generically entails the appearance of massive $U(1)$ factors, whose masses originate from the generalized Green--Schwarz mechanism. In the worldsheet CFT description, we have a single massive $U(1)_\mathbf{Q}$ factor, determined by the affine Abelian current $J_\mathsf{R}$~(\ref{ucurr}).

In analogy to how massless gauge multiplet operators ~(\ref{vgauge}) are constructed, we can build the vertex operator associated the massive $U(1)_\mathbf{Q}$ gauge field in the following way: we tensor on the left--moving side the space--time contribution $\gps_\mathsf{L\,-1/2}^\gm|0\rangle_\textsc{ns}$ with the descendent of the tensor product of an anti--chiral primary of $SU(2)_k/U(1)$ with a chiral primary of $\slr_{k'}/U(1)$, with 
spins identified $j=J$. Such left--moving states have total conformal weight:
\equ{
\Delta = \frac12 + \frac12 + \frac{j}{k} + \frac{J}{k'} = 1 + \frac{2j}{k} \,. 
}
The right--moving part of the $U(1)_\mathbf{Q}$ vertex  operator is identical to the operator $ \mathcal{V}_\mathsf{R}^{(2)}$~(\ref{V2})
seen previously. Upon identifying $j=J$, its total conformal weight is:
\equ{
 \bar \Delta = 1+ \frac12 \mathbf{P}_\text{sh}^2 +\frac{2j}{k} + \Big(\frac{k-2}{4}-j\Big)\gg  \,.
}
Level matching then leads to the equation:
\equ{
 \mathbf{P}_\text{sh}^2 + \Big(\frac{k-2}{2}-2j\Big)\gg =0  \,.
}
We require the $\slr$ spin to be identified as $J=j$ and to satisfy condition~(\ref{Mcond}) and moreover  the $SU(2)$ spin $j$ to lie in the allowed range. It can then be shown that the only state corresponding to an Abelian gauge field
 must be in the untwisted NS sector, with fermions characterized by the charge vector $\mathbf{P}_\text{sh}=(0^{16})$
and hence in the discrete $\slr_{k'+2}/U(1)$ representation $(J,M)=(1,0)$. This is consistent with condition~(\ref{spinM2}) for the operator $\mathcal{V}_\mathsf{R}^{(2)}$. The complete vertex operator describing the massive $U(1)_\mathbf{Q}$ then reads:
\equ{\label{opU1}
\mathcal{V}_{U(1)_\mathbf{Q}} =\psi^\mu_\mathsf{L} \big( G_{-1/2} \,A_{\mathsf{L}\, 1} \, C'_{\mathsf{L}\, 1}\big) \otimes 
\mathbf{Q}\cdot \bar{\partial} \mathbf{X}_\mathsf{R}\,
\ee^{-\sqrt{\frac{2}{\alpha'k'}} \varrho }\,
V_{\mathsf{R}\, 1,m}\,.
}
This state has $J>0$ so, unlike the unbroken gauge group multiplets, is normalizable. Finally the mass of the $U(1)_\mathbf{Q}$ can be determined from its conformal weights $\Delta=\bar\Delta= 1+\frac{2}{k}$ to be:
\equ{
 \mathfrak{m}  = 2 \sqrt{\frac{2}{\alpha'k}} = \frac{4}{\sqrt{\alpha'(\mathbf{Q}^2 -4)}}~. 
}

\section{Comparison between CFT  and supergravity results}
\label{sc:compar}

% what is the section about
%
In this section we compare the spectra computed in the supergravity approximation of Section~\ref{sc:models} with that of the CFT computations in Section~\ref{sc:CFTwarpedEH}. We will see that the spectra indeed agree provided we conjecture a correspondence between the untwisted and twisted sectors of the warped Eguchi--Hanson CFT.

\subsection{Matching the charge supergravity and untwisted CFT spectra}

% agreement with untwisted CFT spectra 
%
Although we are dealing with local heterotic models and not full--fledged compactifications, the spectra of massless states should for consistency be gauge anomaly free. For non--compact warped backgrounds, as the ones under scrutiny in this paper, we have shown in Section~\ref{sc:FiveBranes} that the presence of five--brane charge $\mathcal{Q}_5$ induces the right content of extra (anti--)five--brane (anti--)hyper multiplet states to cancel all irreducible gauge anomalies. We thus expect the spectrum of massless states computed in the CFT to concur with the supergravity results of Table~\ref{tb:pertspectrum}. We observe that we have a precise matching between the untwisted localized CFT spectra of Table~\ref{tab:untwsec} and the hyper multiplet spectra in Table~\ref{tb:pertspectrum} up to two minor caveats: In the untwisted CFT spectra an extra $-\sfrac14$ correction to multiplicities has to be taking into account for vector bundles with some odd entries, see Section~\ref{sec:cftmodels}. Secondly, 
there is a single universal correction $-\sfrac{1}{12}\, c_2$ in the supergravity spectra of Section~\ref{sc:spectra}.

% Higgsing 
%
For Eguchi--Hanson models without warping, i.e.\ $\mathcal{Q}_5=0$, we recall that in Section~\ref{sc:anomalies} this $-\sfrac{1}{12}\, c_2$ correction could be interpreted as a Higgsing process: Upon blowing up the $\ci^2/\zi_2$ singularity two bulk gauginos acquire mass by pairing up with twisted hyper multiplet states. It is not unreasonable to assume the same interpretation of the correction $-\sfrac 1{12}\, c_2$ for warped Eguchi--Hanson models with non--vanishing five--brane charge. However, as stressed in Section~\ref{sec:gaugemu}, because of the non--compactness of the $ \slr/U(1)$ factor the six dimensional gauge fields correspond to non--normalizable vertex operators and hence are not part of the spectrum. Consequently, the CFT multiplicities for hyper multiplets, determined by the indices in~(\ref{ind1}),~(\ref{ind2}) and~(\ref{ind3}), are simply blind to such fractional corrections due to such non--normalizable states.

% Relation between untwisted and twisted states 
%
The situation of the matching between the spectra of the supergravity and the CFT, obtained so far, can be summarized as follows: we have seen that the untwisted states of the CFT given in Table~\ref{tab:untwsec} correctly reproduce the anomaly free heterotic supergravity spectra of Table~\ref{tb:pertspectrum}, taking the universal $-\sfrac{1}{12}\, c_2$ correction into account. Hence, either the twisted states of Table~\ref{tab:twsec} are all massive or they are somehow related to the untwisted sector. Based on~\cite{Giveon:2001up,Aharony:2004xn} we have reasons to believe that marginal deformations of the worldsheet Lagrangian~(\ref{WSactionCFT}) by vertex operators listed in Table~\ref{tab:untwsec} and~\ref{tab:twsec} have couplings which are not independent. 

% what is in the remainder of this section 
%
In the next Subsection we recall the results of~\cite{Giveon:2001up} on two singlet marginal deformations of $\slr/U(1)$ to show that the corresponding twisted and untwisted states in the CFT represent the same uncharged hyper multiplet in supergravity. In Subsection~\ref{sc:conjecture} we observe that untwisted and twisted operators correspond to perturbative versus non--perturbative states  in the $\alpha'$--expansion in  the supergravity limit. Finally, in Subsection~\ref{sc:AdSCFT}  we generalize the argument of Subsection~\ref{sc:deformations}  to an $SU(2) \times \slr$ coset theory. In particular, we use the holographic correspondence in~\cite{Aharony:2004xn} to give arguments in favor of how the relation between couplings of marginal operators should apply to all twisted and untwisted states of the warped Eguchi--Hanson CFT in the double scaling limit.

\subsection{Cigar and Liouville CFT deformations}
\label{sc:deformations} 

% untwisted and twisted deformations
%
Consider the asymmetric current--current deformation and the $(2,0)$ Liouville interaction generated by the untwisted (Table~\ref{tab:untwsec}) and twisted (Table~\ref{tab:twsec}) operators in the singlet representation of the unbroken gauge group, 
\begin{eqnarray} 
\delta S_\text{cur--cur} &=&  \mu_{1,0}  \int  \di^2 z \,\big(J_\mathsf{L}^3+ :\! \psi^\varrho_\mathsf{L} \psi^3_\mathsf{L}\!:\big)\,\mathbf{Q}\cdot \bar\partial \mathbf{X}_\mathsf{R}\,  \ee^{-\sqrt{\frac{2}{\alpha'k}} \varrho} + c.c. \,,  \label{dS} \\
\delta S_\text{Liouv.} & =&  \mu_{\frac{k}{2},\frac{k+2}{2}}  \int \di^2 z\, \big( \psi^\varrho_\mathsf{L} + i\psi^3_\mathsf{L} \big) \, \ee^{-\sqrt{\frac{k}{2\alpha'}}(\varrho + iY_\mathsf{L}) -\frac i2 
 \mathbf{Q}\cdot \mathbf{X}_\mathsf{R}} +c.c.~, 
 \label{defLiouv}
\end{eqnarray}
respectively.~\footnote{For simplicity we give these operators in the 0--picture, thereby omitting the bosonized ghosts. We also leave out the $\er^{5,1}$ contribution.}
The relation between the right--moving $SU(2)_{k}$ supercurrent and the chiral boson $Y_\mathsf{L} $ is given by:
\begin{equation} \label{bil}
 J^3_\mathsf{L} = i\sqrt{\tfrac{k}{\alpha'}}\, \partial Y_\mathsf{L} \,.
\end{equation}
The first operator corresponds to the volume of the resolved two--cycle of the Eguchi--Hanson space, while the second operator gives the SCFT description of the blow--up mode. Consequently, the two deformations~(\ref{dS}) and~(\ref{defLiouv}) cannot be turned on independently; their couplings are related.

% relating marginal defs by Giveon 
%
The exact relation between these two coulings has been determined by~\cite{Giveon:2001up} for the  $\slr_k/U(1)$  CFT with $(2,2)$--worldsheet supersymmetry, which underlies the cigar background geometry. This CFT gives therefore a microscopic description of a two--dimensional Euclidean black hole in type--II string theory.
By looking at the conformal weights of chiral primaries of the $\slr_k/U(1)$ SCFT in Table~\ref{tb:ConfWeights} we see that
the only marginal deformations in this case are the cigar and the $(2,2)$ Liouville deformations, corresponding to~(\ref{dS}) and~(\ref{defLiouv}), respectively. These now read:
\begin{eqnarray} 
\delta S_\text{cigar} &=&  \mu_{1,0}  \int  \di^2 z \, \partial X_\mathsf{L} \,\bar\partial X_\mathsf{R}  \, \ee^{-\sqrt{\frac{2}{\alpha'k}} \varrho} + c.c. \,,  \label{dS2} \\
\delta S_\text{Liouv.} & =&  \mu_{\frac{k}{2},\frac{k+2}{2}}  \int \di^2 z\, \big( \psi^\varrho_\mathsf{L} + i\psi^{X}_\mathsf{L} \big)\big( \psi^\varrho_\mathsf{R} - i\psi^{X}_\mathsf{R} \big) \, \ee^{-\sqrt{\frac{k}{2\alpha'}}(\varrho + i(X_\mathsf{L}-X_\mathsf{R})} +c.c.  \,,
 \label{defLiouv2}
\end{eqnarray}
where $X= X_\mathsf{L}+X_\mathsf{R}$ is the $\slr_k/U(1)$ compact boson parametrizing the circle coordinate of the cigar. 
A two--point function calculation shows that the couplings of these two interactions are related through~\cite{Giveon:2001up}:
\equ{  
\mu_{1,0} = C_{k} \,\big(\mu_{\frac{k}{2},\frac{k+2}{2}} \big)^{\frac 2k} \,, \qquad  \text{with}\;  C_{k} =  \left(-\frac{\pi}{k}\right)^{\frac2k} \frac{\Gamma\left( \frac{k-1}{k}\right)}{\pi\, \Gamma\left( \frac{1}{k}\right)}~.
}
%

% Extending Giveon result to our case
%
This result, although only established for the  $\slr_k/U(1)$ model, holds for~(\ref{dS}) and~(\ref{defLiouv}) as well.
This confirms that in the heterotic warped Eguchi--Hanson theory the two singlet operators leading to~(\ref{dS}) and~(\ref{defLiouv}) do not correspond to different massless bosons in the low--energy theory.  Thus, the untwisted singlet in Table~\ref{tab:untwsec} gives the CFT description of the blow--up modulus in supergravity, the relation being the following (see~\cite{Giv}):
\equ{\label{corrmu}
\left |\frac{\mu_{1,0}}{C_k} \right| = \left(\frac{g_s}{\lambda}\right)^2 = \frac{a^2}{\alpha'}~.
}
Since the Liouville potential encodes worldsheet instanton corrections to the resolved background geometry,  the twisted singlet state in Table~\ref{tab:twsec} gives in contrast an expression for its non--perturbative $\alpha'$--completion. This will be study in more detail in the next Subsection.

%%%%%%%%%%%%%%%%%%%%%

\subsection{Twisted and untwisted vertex operators as perturbative or non--perturbative states}
\label{sc:conjecture}

% Formulate our conjecture 
%
Here, we comment on the behaviour in the supergravity limit of twisted and untwisted vertex operators corresponding to the spectra of Tables~\ref{tab:untwsec} and~\ref{tab:twsec}: 
In Section~\ref{sc:general} we recalled that the conformal factor~(\ref{metric2}) and dilaton~(\ref{solH}) receive $1/\mathcal{Q}_5$ corrections in the finite five--brane charge limit. This translates on the CFT side to $1/k$ corrections to the background metric~(\ref{WSactionCFT}) and dilaton~(\ref{dilCFT}), which can be computed by a method similar to the one used in~\cite{Johnson:2004zq} for light--like asymmetric gaugings of WZW models. As the CFT at level $k$ is related to the curvature of the background geometry, this $1/k$ expansion corresponds to a (higher derivative) $\alpha'$--expansion in supergravity. As can be seen from the multiplicities in Tables~\ref{tab:untwsec} combined with~\ref{tab:twsec} and spin values~\eqref{p2_sim_k} we have:
\equ{\label{pert-npert}
\renewcommand{\arraystretch}{1.5} 
\begin{array}{c|c|c}
%\hline 
\text{sector} & \text{dynamical  marginal deformation} & \text{scaling of}~J~\text{for large}~k 
\\\hline %\hline 
\text{untwisted} &   \text{perturbative} & J \lesssim \mathcal{O}(\sqrt k) 
\\% \hline 
 \text{twisted} & \text{non--perturbative} &  J \sim \mathcal{O}(k)
 \\  %\hline 
  \end{array}
}
Hence, the normalizable marginal deformations involving the vertex operators~(\ref{V1}) and~(\ref{V2}) can be classified as perturbative or non--perturbative depending on their $\slr/U(1)$ spin: As all these operators are in discrete $\slr/U(1)$ representations, they are localized and therefore generically contain a factor $\exp(-\sqrt{2/\ga' k} J \varrho)$. Consequently, the vertex operators with $J \sim k$ are exponentially suppressed in the supergravity limit,  $k \rightarrow \infty$, while for  $J\sim\sqrt{k}$ the factors of $\sqrt{k}$ in the exponential cancel and the corresponding operators are not suppressed. 

Thus all massless bosonic states of Table~\ref{tab:untwsec}, which correspond to untwisted marginal vertex operators, are 
perturbative, while all the twisted states of Table~\ref{tab:twsec}, which are suppressed in the large $k$ limit, are non--perturbative in the supergravity limit.

%%%%%%%%%%%%%%%%%%%%

\subsection{Relating twisted and untwisted massless states through holography: a conjecture}
\label{sc:AdSCFT} 

% the conjecture 
%
In this final Subsection we propose a conjecture that relates massless CFT states in the twisted sector (Table~\ref{tab:twsec}) with the untwisted sector (Table~\ref{tab:untwsec}), by generalizing the conclusion of Subsection~\ref{sc:conjecture} to the couplings of their corresponding
marginal deformations.

For this purpose, we consider a four dimensional CFT closely related to the warped Eguchi--Hanson CFT~(\ref{cftback}), namely the symmetric gauging of $\slr_k \times SU(2)_k$ giving the $(4,4)$ CFT:
\equ{\label{nscir}
 \er^{5,1} \times \left( \frac{\slr_k}{U(1)} \times \frac{SU(2)_k}{U(1)}\right) / \zi_k~, 
}
studied in~\cite{Aharony:2004xn}. This CFT was shown to provide a T--dual description of the near--horizon geometry of $k$ NS--five--branes evenly distributed on a circle of radius $a$. 

% holographic correspondence 
%
It has been proposed in~\cite{Aharony:2004xn} that in the double--scaling limit there exists a holographic correspondence between the string theory describing the background~(\ref{nscir}) and the little string theory (LST), i.e.\ the non--gravitational theory, living on 
the world--volume of the five--branes. In particular, in the asymptotic limit where $\varrho \rightarrow \infty$, these authors have conjectured a duality, similar to the AdS/CFT correspondence, between vertex operators in the CFT and multi--trace operators in the dual LST. 

In the asymptotic limit the CFT operators for the asymmetrically gauged WZW model~(\ref{nscir}) can be expressed in terms of vertex operators of the CHS theory $ \er^{5,1}  \times \er_\mathcal{Q} \times SU(2)_k$ modded out by a discrete symmetry. On the LST side 
the operators of the low--energy gauge theory are built from scalars in the adjoint representation of $SU(k)$, denoted by $\Phi^i$, $i=6,..,9$, which result from promoting the four transverse coordinates of the $k$ NS--five--brane to $SU(k)$ matrices.

% holographic dictionary 
%
The holographic dictionary established in~\cite{Aharony:2004xn} thus relates normalizable vertex operators corresponding to discrete $\slr/U(1)$ representations to multi--trace operators in the LST in a similar spirit as the AdS/CFT correspondence:
\equ{\label{dic}
\ee^{ip_\mu x^\mu}\ee^{-\varphi_\mathsf{L}-\varphi_\mathsf{R}} \big( \psi_\mathsf{L} \psi_\mathsf{R} V_j)_{j+1;m,\bar m} 
\ee^{-\sqrt{\frac{2}{\alpha' k}}(j+1)\varrho}
\qquad \longleftrightarrow \qquad 
  \widetilde{\text{tr}} \big(\Phi^{(i_1} \Phi^{i_2}  \cdots \Phi^{i_{2j+2})} \big)~,
}
where $ \widetilde{\text{tr}}$ denotes multi--traces that are symmetric and traceless in the $(i_1,\ldots,i_{2j+2})$ indices.  In addition,
$\varphi_{\mathsf{L},\mathsf{R}}$ denote the left-- and right--moving bosonized super-conformal ghost system. Also
$\psi_{\mathsf{L},\mathsf{R}}^a$, $a=3,\pm$ represent the left-- and right--moving fermions of the supersymmetric $SU(2)_k$, which
combine with a bosonic $SU(2)_{k-2}$ affine primary $V_j$ into a supersymmetric primary $\big( \psi_\mathsf{L} \psi_\mathsf{R} V_j\big)_{j+1;m,\bar m}$ of total spin and charges $(j+1;m,\bar m)$. 

Perturbing the worldsheet action $S_0$ of the CFT defined in~(\ref{nscir}) by a marginal deformation involving a primary operator~(\ref{dic}), as
\equ{
S_0 + \mu_{j;m,\bar m } \int \di^2 z\,\ee^{ ip_{\mu} x^{\mu} }\, \ee^{-\varphi_\mathsf{L}-\varphi_\mathsf{R}}  \, G_{-\frac12} \bar G_{-\frac12}\,\big( \psi_\mathsf{L} \psi_\mathsf{R} V_j)_{j+1;m,\bar m}  
\ee^{-\sqrt{\frac{2}{\alpha' k}}(j+1)\varrho}~, 
\label{couplgen}
}
results in a displacement of the five--branes away from their initial circular configuration in some particular direction, and thus corresponds to turning on a VEV for the gauge theory operator in~(\ref{dic}). In particular, this breaks the $SU(k)$ gauge symmetry to $U(1)^{k-1}$. Since the moduli space of the six-dimensional LST is determined by the position of the $k$ \textsc{ns} five-branes in the four transverse dimensions, the VEVs $\langle \widetilde{\text{tr}} (\Phi^{(i_1} \Phi^{i_2}  \cdots \Phi^{i_{2j+2})} \big) \rangle$ are in general related to one another through the moduli of the LST. Using the dictionary~(\ref{dic}) then shows that the couplings $\mu_{j;m,\bar m }$ in~(\ref{couplgen}) are related likewise, with~(\ref{corrmu}) being a particular case of this dependence between couplings of marginal deformations. Since the holographic dictonary~(\ref{dic}) involves multi--trace operators in the LST, for general couplings $\mu_{j;m,\bar m }$ the correspondence is not necessarily one--to--one.

% extension to our case 
%
Even though the LST dual to the warped Eguchi--Hanson heterotic CFT~(\ref{cftback}) is unknown, we believe that a holographic correspondence similar to~(\ref{dic}) should exist: Ref.~\cite{Aharony:2004xn} gives evidence that such a duality can be verified for various known examples of CFTs underlying NS--five--brane backgrounds that asymptote to a linear dilaton theory, which is the case for~(\ref{cftback}). 
In this perspective, we propose that the untwisted and the twisted massless states of Section~\ref{sc:MasslessCFTspec} are related
to one another through the moduli of the dual low--energy gauge theory in the fashion just explained.
As emphasized, this correspondence need not be one--to--one, which could account for the mismatch between the multiplicities of the untwisted and twisted states in Tables~\ref{tab:untwsec} and~\ref{tab:twsec}, respectively. Furthermore, in light of Subsection~\ref{sc:conjecture}  we propose that the hyper multiplets in Table~\ref{tb:pertspectrum} are described in the CFT and their multiplicities given by the untwisted states in Table~\ref{tab:untwsec}, while the twisted states in Table~\ref{tab:twsec} represent their non-perturbative in $\alpha'$ completion.

Verifying this conjecture using holography  is unfortunately beyond the scope of this paper, since this would require uncovering first the gauge theory dual to the CFT given in~(\ref{cftback}). Alternatively, one might be able to prove this conjecture  by computing and comparing $n$--point functions with insertions of the relevant marginal vertex operators, and thus understand the relations between the multiplicities of twisted and untwisted states. That however would be a formidable task for the CFT under consideration, with its large and complicated set of marginal deformations.

\newpage 
\section{Compact $\boldsymbol{T^4/\Intr_2}$ resolution models}
\label{sc:compactmodels}

In this section we apply our analysis of non--compact warped Eguchi--Hanson spaces to the compact orbifold $T^4/\Intr_2$ and its smooth resolution. To this end we first give a description of the intersection ring of the Divisors of the resolved $T^4/\Intr_2$. Using this topological data we compute the spectrum on such K3 resolutions.

\subsection{Global description of the compact $\boldsymbol{T^4/\mathbbm{Z}_2}$ orbifold resolution}

Following the methods described in the works \cite{Denef:2004dm,Lust:2006zh} (see also \cite{Nibbelink:2009sp,Blaszczyk:2010db}) we can extend the description of the local resolution of a single $\mathbbm{C}^2/\mathbbm{Z}_2$ singularity to a global description of the resolution of the compact orbifold $T^4/\Intr_2$. Since this orbifold has 16 $\Cplx^2/\Intr_2$ singularities, one introduces $4+4$ ordinary divisors $D_{1,n_1}$ and $D_{2,n_2}$ with $n_1,n_2=1,2,3,4$. Each pair of $(D_{1,n_1}, D_{2,n_2})$ intersects at one of these 16 fixed points. In blow--up each of them is replaced by an exceptional divisor $E_{n_1n_2}$. They intersect with the ordinary divisors dictated by the local $\Cplx^2/\Intr_2$ resolution description, i.e.\ 
\equ{ 
D_{1,n_1} E_{n_1'n_2'} = \gd_{n_1n_1'}~, 
\qquad
D_{2,n_2} E_{n_1'n_2'} = \gd_{n_2n_2'}~. 
}
In addition the four--torus has two--torus subspaces. In blow--up they lead to so--called inherited divisors $R_1$ and $R_2$ subject to the linear equivalence relations
\equ{
D_{1,n_1} = \frac 12 \Big( R_1 - \sum_{n_2} E_{n_1n_2} \Big)~, 
\qquad 
D_{2,n_2} = \frac 12 \Big( R_2 - \sum_{n_1} E_{n_1n_2} \Big)~.
\label{linequivT4Z2}
}
Their non--vanishing basic intersections are 
\equ{
R_1 D_{2,n_2} = R_2 D_{1,n_1} = 1~, 
\qquad 
R_1 R_2 = 2~. 
}
The linear equivalence tell us that a basis of $H^2$ is provided by $R_1,R_2$ and $E_{n_1n_2}$. Given the intersection numbers given above their non--vanishing (self--)intersection read
\equ{
R_1R_2 = 2~, 
\qquad 
E_{n_1n_2} E_{n_1'n_2'} = -2\, \delta_{n_1n_1'} \delta_{n_2n_2'}~.
}
Using the splitting principle and the linear equivalences one finds that for the resolution of $T^4/\Intr_2$ we get 
\equ{ 
c_1 \sim 0~, 
\qquad 
c_2 \sim - \frac 34 \sum_{n_1,n_2} E_{n_1n_2}^2~. 
}
Inserting the self--intersection numbers we find that 
$c_2 = -\frac 34 * 16 * (-2) = 24$, as expected for a K3 surface \cite{Nahm:1999ps}.

\subsection{Model building on compact resolutions}

In addition to describe models on the resolution of $T^4/\Intr_2$ we also have to specify the gauge background. Just like the non--compact case we take a combination of line bundles:
\equ{ 
\frac{\mathcal{F}}{2\pi} = - \frac 12\, 
\sum_{n_1,n_2, I} E_{n_1n_2}\,  Q_{n_1n_2\, I} \, \textsf{H}^I~,
}
in terms of a set of line bundle vectors $\mathbf{Q}_{n_1n_2}$ labeled by the fixed point indices $(n_1,n_2)$. In this case we find for the non--integrated Bianchi identity 
\equ{
\frac 1{4\pi^2 \ga'}
 \d \mathcal{H}  = 
 \sum_{n_1,n_2} 
 \Big( \frac 14\, \mathbf{Q}_{n_1n_2}^2  - \frac 32 \Big) 
 E_{n_1n_2}^2~. 
 \label{BInonint}
}
When we insert the self--intersections we can express this as 
\equ{ 
-\mathcal{Q}_5 = 
24 - \frac 12\, \sum_{n_1,n_2} \mathbf{Q}_{n_1n_2}^2~.
}
Since we here describe a compact model, $\mathcal{Q}_5 > 0$ means we need anti--five--branes and consequently six--dimensional supersymmetry is broken. In the case that $\tN = - \mathcal{Q}_5/2 > 0$ we know that we need $\tN$ NS five--branes to cancel the anomalies.

It is also possible that $\mathcal{Q}_5 = 0$, while the non--integrate Bianchi identity \eqref{BInonint} is non--vanishing. This means that locally near each of the resolved singularities one would infer that there are either five--branes or anti--five--branes, but their charges in total precisely compensate. A concrete example of such a configuration was presented in \cite{Honecker:2006qz,Nibbelink:2007rd}: In this model one has the bundle vectors 
\equ{
\mathcal{Q}_{1n_2} = (-1^2, 0^{14})~,
\qquad 
\mathcal{Q}_{2n_2} = \mathcal{Q}_{3n_2} = (1^2, 0^{14})~,
\qquad 
\mathcal{Q}_{1n_2} = (3^2, 0^{14})~. 
}
Hence, at each of the 12 fixed points, $(n_1,n_2)$ with $n_1=1,2,3$ and $n_2=1,2,3,4$, one needs a single five--brane. While at each of the four remaining fixed points, $(4,n_2)$ with $n_2=1,2,3,4$, there are three anti--five branes present. Clearly their contributions exactly cancel out: This means that the anti--hyper multiplets from the anti--five--branes have paired up with corresponding hyper multiplet states and decouple from the massless spectrum. Consequently, the combined perturbative spectrum, given by 
\equ{
10 \cdot (\rep{2},\rep{28}) + 46\cdot (\rep{1},\rep{1})
}
w.r.t.\ the unbroken gauge group $SU(2)\times SO(28)$, is anomaly free. Up to one additional singlet $(\rep{1},\rep{1})$ this is the spectrum of the model using the standard embedding, i.e.\ where the gauge connection is set equal to the spin connection (see e.g.\ \cite{Nibbelink:2007rd}).

\newpage 
\section{Conclusions}
\label{sc:concl}

% recall context of the paper 
%
We considered the heterotic string on the warped Eguchi--Hanson space in the presence of line bundle gauge backgrounds. We reviewed and compared two alternative descriptions of such heterotic backgrounds: in the leading order $\alpha'$--expansion we employed supergravity techniques. In the double scaling limit, in which both the blow--up radius $a$ and the string coupling $g_s$ tend to zero simultaneously, we exploited an exact CFT description. The underlying worldsheet description was built on an asymmetric $U(1)_\mathsf{L} \times U(1)_\mathsf{R}$ gauging of the WZW model on the group space $SU(2)_k \times SL(2,\Real)_{k'}$. In particular, by studying the residual gauge invariance in the blow down limit we were able to identify the origin of the $\zi_2$ orbifold of the Eguchi--Hanson space from the perspective of continuous $\slr/U(1)$ representations in the CFT. This completes an argument in~\cite{Carlevaro:2008qf} which relied on the existence of a Liouville potential in the spectrum of discrete representations.

% general spectra computation
%
For both regimes there are methods at our disposal to compute the massless spectrum of the heterotic string on these backgrounds: In the supergravity approximation we used a multiplicity operator (which may be thought of as a representation dependent index) to determine the spectra. In the double scaling limit we determined the massless states by classifying all the marginal operators that the CFT admits. In either regime we computed the spectra in a systematic way for a large class of models simultaneously, by concentrating on a general line bundle vector $\mathbf{Q} = (p_1^{N_1},\ldots, p_n^{N_n}, 0^N)$ with $N_1 + \ldots + N_n + N = 16$ where $p_i$ are some different positive integers. 

% sugra spectra
%
The spectra in the supergravity approximation can be read off from Table \ref{tb:pertspectrum}. As this table shows the multiplicities are not necessarily integral, which is an artifact due to the non--compactness of the (warped) Eguchi--Hanson space. This effect has been observed for various other non--compact resolutions in the literature~\cite{Nibbelink:2007rd,Nibbelink:2007pn,Nibbelink:2008qf}. 

% CFT spectra
%
The second way to determine the spectra on the warped Eguchi--Hanson space takes the full (anti--)five--brane backreacted geometry into account. Given that the CFT in the double scaling limit contains a $\Intr_2$ orbifold acting on its $SU(2)_k$ contribution, the CFT spectra fall into two classes. The corresponding untwisted and twisted massless states are collected in Tables~\ref{tab:untwsec} and~\ref{tab:twsec}, respectively. By comparing the results  for supergravity and CFT spectra, it follows immediately that untwisted spectra in Table~\ref{tab:untwsec} is essentially identical to the supergravity spectra collected in Table~\ref{tb:pertspectrum}. There is a difference in the fine print of the spectra calculated in either regime: The untwisted CFT and the supergravity spectra differ up to a universal fractional correction, which is due to a Higgsing process. The CFT multiplicity index appears to be blind to this effect. 

Furthermore, to interpret the apparent redundancy of the twisted and untwisted CFT spectra as compared to the supergravity results,
we propose that the couplings of the corresponding marginal deformations of the coset CFT are related: 
The untwisted and twisted massless CFT states provide   
perturbative and non--perturbative $\alpha'$--descriptions of the same hyper multiplets in the supergravity regime.
To support our conjecture, we rely on the proposed duality of Ref.~\cite{Aharony:2004xn} relating an asymptotically linear dilaton CFT to a little string theory living on a configuration of NS--five--branes. 

Proving this conjecture is unfortunately beyond the scope of this work because the low--energy gauge theory dual to the warped Eguchi--Hanson CFT is unknown. Of course, the perspective of uncovering a heterotic AdS/CFT type correspondence for the warped Eguchi--Hanson with line bundle and five--brane flux is extremely interesting. In this case the dual low--energy gauge theory would likely be given by a type of little string theory. However, in the absence of such a low--energy description one could verify this proposal by computing and comparing the $n$--point functions of the appropriate marginal operator insertions. This would not only be instrumental in reconciling the difference between the combined twisted/untwisted CFT and supergravity spectra, but would also improve the understanding of the apparent mismatch between the multiplicities of untwisted and twisted CFT states.

% five-branes and anomaly cancellation 
%
When the five--brane charge is non--vanishing, these untwisted CFT or supergravity spectra appear to be anomalous. However, following the logic of Refs. \cite{Aldazabal:1997wi,Honecker:2006dt} we showed that these spectra are free of irreducible gauge anomalies provided that five--brane states are included. Given that the chirality in six dimensions of gauginos and hyperinos is fixed by supersymmetry, it turns out that it is simply impossible to cancel the anomalies without breaking supersymmetry in most cases: The only way to construct models free of irreducible anomalies is to allow for anti--hyper multiplets  (i.e.\ hyper multiplets with the opposite chirality as that in the perturbative string sector) to live on anti--five--branes. Hence, possibly at the price of supersymmetry breaking, the anomalous perturbative string spectra can be completed by non--perturbative five--brane states to become anomaly free. 

% compact models
%
Since the (warped) Eguchi--Hanson space is non--compact, it is a bit premature to conclude from this analysis that supersymmetry is indeed broken. To see what happens when various locally (warped) Eguchi--Hanson spaces are glued together, we investigate the resolution of the $T^4/\Intr_2$ orbifold in which the local Bianchi identities are never satisfied near the resolved singularities; only the global Bianchi identity is fulfilled. We find that the local five--branes and anti--five--branes precisely compensate each other and the full model is free of anomalies while supersymmetry is preserved.

\subsection*{Acknowledgements}

We would like to thank Dan Israel for extensive discussions and email communications. We would further like to thank Emilian Dudas, James Gray, Gabriele Honecker, Susha Parameswaran, Fabian R\"uhle for valuable discussions at various stages of this project. 
SGN would like to thank the CPTH at the Ecole Polytechnique and LC the ASC at the Ludwig--Maximilians--Universit\"at M\"unchen and the Max--Planck--Instute for kind hospitality. This research has been supported by the "LMUExcellent" Programme.

\appendix 
\def\theequation{\thesection.\arabic{equation}} 

\setcounter{equation}{0}
\section{Some trace identities} 
\label{sc:traces}

This appendix summarizes some trace identities relevant for anomaly investigations in six dimensions. 

% SU(N) traces

The adjoint representation $\rep{Ad}$ of $SU(N)$ is obtained as $\rep{N}\otimes \crep{N} \ominus \rep{1}$ with $\rep{N}$ the fundamental and $\crep{N}$ the anti--fundamental representation and he anti--symmetric tensor representation is defined as $[\rep{N}]_2 = \rep{\tfrac 12 N(N-1)}$. Consequently we have the following trace identities: 
\equa{ 
\text{Tr}_\rep{Ad} F_{SU(N)}^4 & = 
2N \text{Tr}_\rep{N} F_{SU(N)}^4 
+ 6 \big(  \text{Tr}_\rep{N} F_{SU(N)}^2 \big)^2~, 
\\[2ex] 
\text{Tr}_{[\rep{N}]_2} F_{SU(N)}^4 & = 
(N -8) \text{Tr}_\rep{N} F_{SU(N)}^4 
+ 3 \big(  \text{Tr}_\rep{N} F_{SU(N)}^2 \big)^2~. 
}
%

% SO(M) traces

The adjoint representation $\rep{Ad}$ of $SO(M)$ equals the anti--symmetric tensor representation $[\rep{M}]_2$ with $\rep{M}$ the vector representation of $SO(M)$. This leads to 
\equ{ 
\text{Tr}_\rep{Ad} F_{SO(M)}^4  = 
(M -8) \text{Tr}_\rep{M} F_{SO(M)}^4 
+ 3 \big(  \text{Tr}_\rep{M} F_{SO(M)}^2 \big)^2~. 
}
%

% Sp(2\tilde N) traces

The fundamental representation of the symplectic group $Sp(2\tN)$ is denoted by $\rep{2\tN}$. The adjoint representation $\rep{Ad}$ of $Sp(2\tN)$ is its symmetric tensor representation $\rep{\tN(2\tN+1)}$. The irreducible anti--symmetric tensor representation $[\rep{2\tN}]_2 = \rep{\tN(2\tN-1)-1}$ is traceless w.r.t.\ its symplectic form $\eta$. This gives the following: 
\equa{ 
\text{Tr}_\rep{Ad} F_{Sp(2\tN)}^4 & = 
(2 \tN + 8) \text{Tr}_\rep{2\tN} F_{Sp(2\tN)}^4 
+ 3 \big(  \text{Tr}_\rep{2\tN} F_{Sp(2\tN)}^2 \big)^2~, 
\\[2ex] 
\text{Tr}_{[\rep{2\tN}]_2} F_{Sp(2\tN)}^4 & = 
(2 \tN - 8) \text{Tr}_\rep{2\tN} F_{Sp(2\tN)}^4 
+ 3 \big(  \text{Tr}_\rep{2\tN} F_{Sp(2\tN)}^2 \big)^2~. 
}

\setcounter{equation}{0}
\section{Theta functions and characters}
\label{sc:thetachar}

\subsection{Dedekind and higher genus theta functions}

The Dedekind function $\get(\tau)$ is given by 
\equ{ 
\get(\gt) = q^{\frac 1{24}} \prod_{n\geq 1} \big[ 1 - q^n \big]~, 
}
where $q = e^{2\pi i \gt}$ with $\gt$ in the complex upper--half--plain. 
The genus one theta function with characteristics $\ga,\gb$ is defined as 
\begin{equation}
\vartheta[^\alpha_\beta](\gt|\nu) = \sum_n q^{\frac 12 (n + \frac \alpha2)^2} e^{2\pi i  (n + \frac \alpha2) (\nu+\frac\beta2)}~,  
\label{Theta}
\end{equation}
The genus--$n$ theta function with vector valued characteristics $\boldsymbol{\ga} = (\ga_1,\ldots, \ga_n)$ and $\boldsymbol{\gb}=(\gb_1,\ldots,\gb_n)$ is obtained as a product of $n$ genus 
\equ{
\vartheta[^{\boldsymbol{\ga}}_{\boldsymbol{\gb}}](\gt) = \prod_{i=1}^n \theta[^{\ga_i}_{\gb_i}](\gt)~. 
\label{genusnTheta}
}

\boldmath
\subsection{$\widehat{\mathfrak{su}}(2)_{k-2}$ characters}
\label{sc:su2characters}
\unboldmath

As $SU(2)$ characters can be found in standard CFT textbooks, like \cite{DiFrancesco:1997nk,Blumenhagen:2009zz}, we will be brief here. The $\widehat{\mathfrak{su}}(2)_k$ theta functions is given by 
\begin{equation}\label{thSU2}
\Theta_{m,k} (\tau,\nu) = \sum_{n\in\Intr}
q^{k\left(n+\tfrac{m}{2k}\right)^2}
z^{k \left(n+\tfrac{m}{2k}\right)}\,,
\end{equation}
where  $z=\ee^{2\pi i\nu}$ and $m$ can be restricted to lie within the range $m=0, \ldots 2k-1$. In terms of these theta functions the characters $\chi^j$ of the bosonic affine algebra $\widehat{\mathfrak{su}}(2)_{k-2}$ are defined as 
\begin{equation}\label{su2-char}
\chi^j (\nu|\tau) = \frac{\Theta_{2j+1,k} (\nu|\tau)-\Theta_{-2j-1,k} (\nu|\tau)}
{\Theta_{1,2} (\nu|\tau)-\Theta_{-1,2} (\nu|\tau)}\,, 
\end{equation}
for the integral or half--integral  spin in the range $0 \leq j \leq \frac{k-2}2$. By carefully taking the limit $\gn \ra 0$ one finds 
\begin{equation}\label{chi-lim}
\chi^j(\tau) = \chi^j(0|\tau) = \frac {1}{\dsp \get^3(\tau)}\,  q^{\frac{(2j+1)^2}{4k}} \, 
\sum_{n\in\zi} \big(2j+1 + 2k\,n \big)\, q^{n (2j+1 + k\,n)}~,  
\end{equation}
where the expansion is set up such that the powers of $q$ are always positive.

We also mention an identity on $\widehat{\mathfrak{su}}(2)_k$ theta functions, which we use in the present work:
\begin{equation}\label{idSU2}
\Theta_{\nicefrac{m}{p},\nicefrac{k}{p}}(\nu|\tau) = \sum_{n\in\zi_p}\Theta_{m+2kn,pk}\big(\tfrac{\nu}{p}\big|\tau\big)\,.
\end{equation}

\bibliographystyle{paper}
{\small
\bibliography{paper}
}

\end{document}